%% file: main.tex
\documentclass[reprint, superscriptaddress  , amsmath, amssymb, aps]{revtex4-2}
\usepackage{booktabs}
\usepackage{graphicx}
\usepackage{subcaption}
\usepackage[percent]{overpic}
\usepackage{ragged2e}
\usepackage[usenames,dvipsnames]{xcolor}
\usepackage[most]{tcolorbox}
\usepackage{tabularx}
\usepackage{tikz}
\usepackage[normalem]{ulem}
\usepackage{hyperref}
\hypersetup{
    colorlinks=true,
    linkcolor=blue,
    filecolor=red,
    urlcolor=cyan,
    citecolor=red,
    pdftitle={Overcoming cluster algorithm failure in weakly frustrated systems via wavelet
sampling},
    pdfpagemode=FullScreen,
}

\makeatletter
\renewcommand{\@makecaption}[2]{%
  \vskip\abovecaptionskip
  \sbox\@tempboxa{#1: #2}%
  \ifdim \wd\@tempboxa > \hsize
    \justifying #1. #2\par
  \else
    \hbox to\hsize{\hfil\box\@tempboxa\hfil}%
  \fi
  \vskip\belowcaptionskip}
\makeatother

\definecolor{dukeblue}{RGB}{1,33, 105}

\begin{document}
\preprint{APS/123-QED}

\title{Overcoming critical slowing down in frustrated spin systems by learned multiscale sampling. %
}
\author{Gabriele Bandini}
 \email{gbandini@sissa.it}
 \affiliation{SISSA --- International School for Advanced Studies and INFN, via Bonomea 265, 34136 Trieste, Italy}

\author{Giulio Biroli}
\affiliation{Laboratoire de Physique Statistique, École normale supérieure, PSL Research University, 24 rue Lhomond, 75005 Paris, France}

 \author{Patrick Charbonneau}
\affiliation{Department of Physics, Duke University, Durham, North Carolina 27708, USA}
\affiliation{Department of Chemistry, Duke University, Durham, North Carolina 27708, USA}

\author{Andrea Gambassi}
 \affiliation{SISSA --- International School for Advanced Studies and INFN, via Bonomea 265, 34136 Trieste, Italy}

\date{\today}

\begin{abstract}

Cluster algorithms, such as the Swendsen--Wang and Wolff methods, are among the most successful MCMC methods for mitigating critical slowing down in statistical systems. 
These constructive cluster algorithms, however, fail in the presence of even extremely weak frustration. %
Here, we sidestep this fundamental limitation by learning rather than constructing the relevant clusters. Specifically, we use the wavelet conditional renormalization group (WCRG) sampling method to learn the probability distribution of collective fluctuations of a frustrated two-dimensional soft-spin model. Configurations are then generated recursively from coarse to fine scales by sampling conditional wavelet distributions. The WCRG method reproduces the main statistical properties of the system across different phases, including the local-field distribution and the structure factor. At an Ising-like critical point, the conditional dynamics remains decorrelated within \(\mathcal{O}(1)\) sweeps at each scale, yielding an overall sampling complexity of \(\mathcal{O}(\log_2 L)\), thus making WCRG much more efficient than standard local MCMC methods. These results show that learned multiscale sampling can overcome critical slowing down in frustrated systems for which conventional cluster algorithms fail. By assessing the sampling accuracy of different observables, we also clarify the main tradeoff of the WCRG method: the accuracy of the fast sampling scheme depends on the expressiveness of the energy-based model used to estimate the wavelet conditional distributions.
\end{abstract}
\maketitle


\section{Introduction}
\label{sec:intro}

Efficiently sampling high-dimensional probability distributions is a central challenge in computational statistical physics and for many other applications of Markov chain Monte Carlo (MCMC) methods. The difficulty becomes especially acute close to a continuous phase transition, where the correlation length grows with the system size and the local-update dynamics is limited by the corresponding 
critical slowing down~\cite{RevModPhys.49.435,Sokal1997}. 
The equilibration and decorrelation times $\tau$ then scale as
$\tau\sim L^z$,
where \(L\) is the linear system size and \(z\) is a dynamical critical exponent that depends on the microscopic model and on the update rule. For two-dimensional systems, local algorithms generally have \(z\) of order two and can grow up to six in models with disorder~\cite{Agrawal2023,Agrawal2024}.

Cluster algorithms provide one of the most successful strategies for mitigating this problem. In unfrustrated spin systems, the Swendsen--Wang and Wolff algorithms update correlated degrees of freedom collectively and strongly reduce critical slowing down~\cite{landau2021guide}. Various generalizations of this scheme have since been proposed, but for---even weakly---frustrated models comparable improvements have been elusive~\cite{miranda2025percolationcriticalitysystemscompeting,Alfaro2026}. %
Recent works \cite{Zheng2022, miranda2025percolationcriticalitysystemscompeting} have identified the fundamental origin of this failure.
In short, although clusters underlying efficient sampling of the probability in configuration space likely exist, they cannot be identified by any constructive procedure.
This fundamental obstacle calls for a completely new approach to sampling. Could these clusters be (machine) learned instead of constructed? 
Standard, off-the-shelf approaches offer only limited grounds for hope. 
Machine-learning-based approaches have recently emerged as a promising route to accelerating sampling more broadly and have therefore attracted growing interest in statistical physics. Examples include normalizing-flow-augmented Monte Carlo schemes
\cite{Gabrie2022}, autoregressive variational networks \cite{Wu2019},
diffusion-based samplers \cite{Ghio2024,DelBono2026}, and
machine-learning-assisted Monte Carlo algorithms \cite{DelBono2025}. These approaches
show that learned generative models can accelerate sampling in favorable cases,
but they also make clear that efficiency and accuracy are not automatic:
sampling difficulties can persist, depending on the structure of the target
distribution, the training procedure, and the expressiveness of the model. This
motivates the consideration of architectures that combine learning with physical structure.

The approach followed here belongs to this broad line of learned sampling
methods, explicitly guided by renormalization-group (RG) ideas from
statistical physics \cite{Kadanoff1966,Wilson1971,WilsonKogut1974}. 
More specifically, the wavelet conditional renormalization group (WCRG) introduced in
Ref.~\cite{Marchand2023},
decomposes the target distribution scale by scale through an orthogonal wavelet transform. This transform separates
a configuration into coarse variables, which encode long-wavelength structure,
and wavelet coefficients, which describe the fluctuations added at each finer
scale. The learned functions are then the various conditional distributions for the wavelet
fluctuations at fixed coarse variables. Each approximation is therefore tied to
a definite length scale and to a specific step of the reconstruction, thus making the
learned model interpretable and providing a natural route for
systematic improvements.

In the present work, the conditional distributions are learned from equilibrium Monte Carlo configurations by score matching, which fits unnormalized probability models by matching gradients of log-probabilities and therefore avoids the direct evaluation of normalization constants \cite{1046920.1088696}. Once the conditional distributions have been inferred, microscopic configurations can be generated recursively: the coarsest field is sampled first, and the microscopic configuration is then reconstructed by conditional Monte Carlo sampling of the wavelet variables at successively finer scales. As discussed in Ref.~\cite{Marchand2023}, this sampling is expected to remain fast because long-range correlations are already encoded in the conditioning coarse fields and do not have to emerge through slow microscopic dynamics.

In this work, we show that 
this strategy is capable of 
bypassing the failure of conventional cluster algorithms in frustrated systems. 
In particular, we consider soft-spin \(\phi^4\) models with competing interactions on a square lattice in two spatial dimensions, the coupling stencil of which encompasses the axial next-nearest-neighbor Ising (ANNNI), biaxial next-nearest-neighbor Ising (BNNNI), and third-nearest-neighbor Ising (3NNNI) geometries. 
Our numerical analysis focuses primarily on the soft-spin BNNNI (sBNNNI) model, %
whose phase diagram contains homogeneous ferromagnetic and paramagnetic regimes separated by an Ising-like critical line as well as modulated disordered and ordered phases.

This analysis leads to three main results.
(i) The generated configurations reproduce the local field distributions and the 
two-point correlations, as measured by the structure factor, across representative regions of the sBNNNI phase diagram. The method captures both homogeneous and finite-wavevector structures, although the strongly constrained antiphase remains a challenging case. 
(ii) At the Ising-like critical point, the autocorrelation time of conventional local Monte Carlo dynamics grows approximately as \(L^{2.16}\), whereas the conditional wavelet updates remain decorrelated in \(\mathcal O(1)\) sweeps at every scale. Because the number of reconstruction levels grows as \(\log_2 L\), the number of conditional sweeps required to generate an independent critical configuration scales overall as \(\mathcal O(\log_2 L)\). 
(iii) For configuration-wide observables, the magnetization distribution is accurately reproduced, 
while the histogram of the microscopic energy displays a residual mismatch. 
We relate this discrepancy to the finite-dimensional ansatz used for the learned conditional energies and to the difference between the original Gibbs measure and the approximate hierarchical measure sampled by the reconstruction procedure.

The rest of the paper is organized as follows. 
In Sec.~\ref{sec:models}, we introduce the frustrated soft-spin models and discuss the phase diagram of the sBNNNI model. In Sec.~\ref{sec:methods}, we present the WCRG decomposition, the learned conditional energies, and the recursive sampling procedure. Section~\ref{sec:numerical-results} compares Monte Carlo configurations with WCRG-generated samples across representative points of the phase diagram. In Sec.~\ref{sec:no-csd}, we analyze the autocorrelation times at criticality and  demonstrate that conditional wavelet updates eliminate critical slowing down.
Section~\ref{sec:config-wise-obs} considers configuration-wide observables, with particular emphasis on the distributions of the magnetization and of the microscopic energy. %
Section~\ref{sec:conclusions} summarize the findings and speculates on possible improvements and applications.

Technical details are provided in the appendices.
Appendix~\ref{app:T0_BNNNI} discusses the zero-temperature analysis of the
sBNNNI model and the origin of the extended IC regime.
Appendix~\ref{app:fss} presents the finite-size scaling analysis used to locate
the Ising-like critical point. Appendix~\ref{app:wavelet-conventions} describes
the implementation of the orthogonal wavelet transform and the conventions used
throughout the paper. Finally, App.~\ref{app:conditional-learning}
contains the details of the parameterizations of the conditional energies, score-matching procedure,
and conditional sampling algorithm used in the
WCRG implementation.
\begin{figure*}[ht!!]
    \centering
    \includegraphics[width=0.92\textwidth]{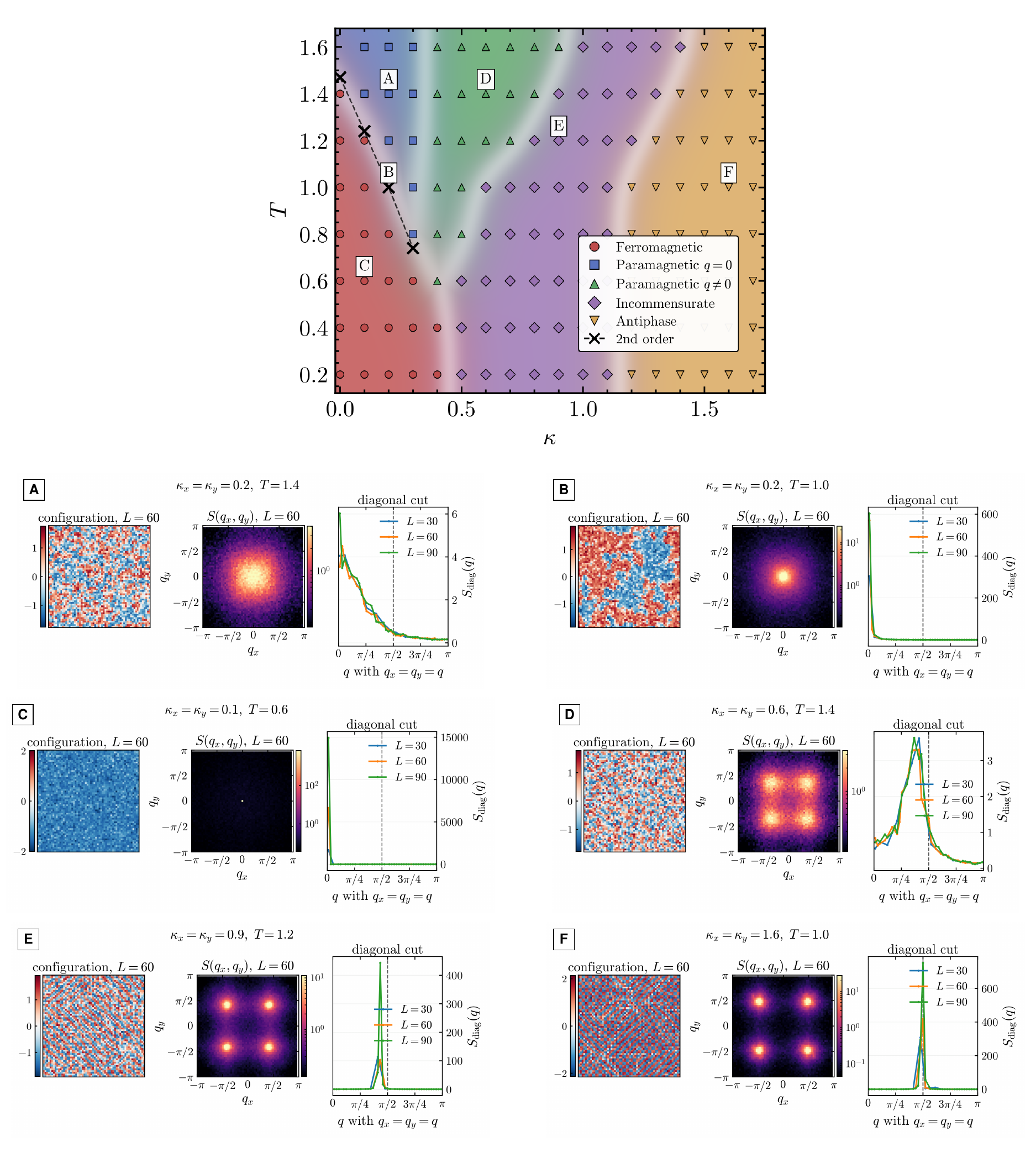}
    \caption{\small  
Phase diagram of the sBNNNI model with $\lambda=1$ and representative behaviors in the various phases. 
\textbf{Top:} Phase diagram in the $(\kappa,T)$ plane obtained from Monte Carlo simulations. The five phases (different colored zones) are identified from the behavior and finite-size scaling of the structure factor $S(q_x,q_y)$ (see Sec.~\ref{sec:models}), sampled at the points indicated by symbols in the phase diagram. 
While the faint boundaries between the various phases are approximate guides to the eyes, the Ising-like ferromagnetic-to-paramagnetic phase transition was quantitatively determined via finite-size scaling of the magnetic susceptibility using Ising critical exponents (black crosses, see App.~\ref{app:fss}). 
Note that the IC phase here extends over a broader region of the phase diagram than in the $\lambda\rightarrow\infty$ model studied in Ref.~\cite{Hu2021_2}, as
further discussed in App.~\ref{app:T0_BNNNI}.
Representative illustrations of the five different phases and the Ising critical line are provided for points A--F. 
\textbf{Bottom:} For each representative point (A--F): a typical equilibrium configuration in real space (left), the averaged
structure factor $S(q_x,q_y)$ (center), and its diagonal cut $S_{\rm diag}(q)$ (see Eq.~\eqref{eq:Sdiag})
for three different system sizes $L$ (right). The peak position and the size scaling of its amplitude distinguishes homogeneous from modulated phases as well as ordered from disordered regimes (see Sec.~\ref{sec:models}).
}
\label{fig:phase-diagram-BNNNI}
\end{figure*}

\section{Models}
\label{sec:models}

In order to leverage wavelet decompositions, which naturally act on real-valued signals \cite{192463, Daubechles+2009+564+652},  we here consider models with continuous spin variables $\phi_0(i) \in \mathbb{R}$, where $i$ denotes the lattice site at position \(\mathbf r_i=(x_i,y_i)\).
The microscopic energy 
on a square lattice is then taken to be 
\begin{equation}
\begin{split}
E_0(\phi_0) = &\frac{1}{2T}\sum_{i,j} \phi_0(i)\, K_{ij}
\phi_0(j) \\
& + \sum_i \Big[\,\phi_0^2(i) + \lambda\big(\phi_0^2(i) - 1\big)^2\Big],
\end{split}
\label{eq:Hsoft}
\end{equation}
Note that, following Refs.~\cite{Marchand2023, doi:10.1142/S0129183116501084}, \(E_0\) is the dimensionless energy entering the Gibbs weight,
\(p_0(\phi_0)\propto e^{-E_0(\phi_0)}\). It does not take the conventional form,
\(H(\phi_0)/T\), with a temperature-independent Hamiltonian \(H\), because the
factor \(1/T\) multiplies only the quadratic interaction kernel, while the local
potential is not rescaled by temperature. As a consequence, the low-temperature  limit requires some care, as discussed in
App.~\ref{app:T0_BNNNI}.
The frustrated quadratic interactions are captured by a $5\times 5$ translationally-invariant matrix
\begin{equation}
\mathbf{K} =
\begin{pmatrix}
0 & 0 & \kappa_y & 0 & 0 \\[4pt]
0 & \kappa_d & -J & \kappa_d & 0 \\[4pt]
\kappa_x & -J & 0 & -J & \kappa_x \\[4pt]
0 & \kappa_d & -J & \kappa_d & 0 \\[4pt]
0 & 0 & \kappa_y & 0 & 0
\end{pmatrix}
\label{eq:stencil}
\end{equation}
with rows and columns encoding the relative displacements $\Delta x, \Delta y \in \{+2,+1,0,-1,-2\}$, i.e.,
$K_{ij} = [\mathbf{K}]_{\Delta x,\Delta y}$, where $\mathbf r_i- \mathbf r_j = (\Delta x,\Delta y)$. 
In practice, nearest-neighbor interactions are characterized by ferromagnetic coupling $J>0$, axial next-nearest neighbors along $\hat{x}$ and $\hat{y}$ by antiferromagnetic couplings $\kappa_x\ge 0$ and $\kappa_y\ge 0$, respectively, and diagonal nearest neighbors by $\kappa_d\ge 0$. All other interactions  vanish. 

The quartic term in Eq.~\eqref{eq:Hsoft} controls spin stiffness. In the Gaussian limit $\lambda\rightarrow0$, $E_0(\phi_0)$ in Eq.~\eqref{eq:Hsoft} reduces to
a free (quadratic) theory; in the hard-spin limit $\lambda\to\infty$, fields are restricted to values $\phi_0(i)=\pm1$, thus recovering Ising spins. For finite $\lambda$, the model is a soft-spin $\phi^4$ theory which, at equilibrium, belongs to 
the \emph{same universality class} as the corresponding frustrated Ising model, differing only at subleading corrections to scaling~\cite{doi:10.1142/S0129183116501084}. %
The various frustrated soft spin models %
mentioned in Sec.~\ref{sec:intro} are then defined as:
(a) ANNNI with $\kappa_x>0$, $\kappa_y=0$, $\kappa_d=0$,
(b) BNNNI with $\kappa_x=\kappa_y > 0$, $\kappa_d=0$, and
(c) 3NNNI with $\kappa_x=\kappa_y >0$, $\kappa_d>0$.
In this work we focus on the soft-spin BNNNI (sBNNNI) model with $\lambda = 1$. Without loss of generality we set $J = 1$ and vary $\kappa = \kappa_x = \kappa_y$ and $T$. 

The resulting equilibrium phase diagram, which we  determined using Monte Carlo simulations (see Fig.~\ref{fig:phase-diagram-BNNNI}), exhibits five distinct phases. The structure factor
\begin{equation}
\label{eq:structure-factor}
S(q_x,q_y)
=
\frac{1}{L^2}
\left\langle
\left|
\sum_i
\phi_0(i)\,
e^{i(q_x x_i + q_y y_i)}
\right|^2
\right\rangle,
\end{equation}
distinguishes these phases from the diagonal cut
\begin{equation}
S_{\rm diag}(q)
=
S(q,q),
\label{eq:Sdiag}
\end{equation}
and the scaling of its peak amplitude with linear system size \(L\) \cite{doi:10.1142/5715}. In particular:
\begin{itemize}
    \item the ferromagnetic (FM) phase exhibits a peak at $q=0$,
the amplitude of which scales 
extensively, i.e., as $S_{\rm diag}(q=0) \sim L^2$ (see Fig.~\ref{fig:phase-diagram-BNNNI}C);
\item the homogeneous paramagnetic (PM) phase exhibits a peak at \(q=0\), the amplitude of which  remains finite as $L\to\infty$ (see Fig.~\ref{fig:phase-diagram-BNNNI}A);
\item the modulated paramagnetic phase (MPM) exhibits a peak at \emph{finite} $q>0$,
the amplitude of which also remains finite as $L\to\infty$ (see Fig.~\ref{fig:phase-diagram-BNNNI}D);
\item the incommensurate (IC) phase presents modulated oder and its structure factor exhibits a peak at finite $0<q<\pi/2$,
the amplitude of which scales as $S_{\rm diag}(q) \sim L^{2-\eta}$ (see Fig.~\ref{fig:phase-diagram-BNNNI}E);
\item the antiphase (AF) corresponds to a checkerboard pattern with a width of two lattice spacings and is characterized by a peak at $q=\pi/2$, the amplitude of which scales extensively 
(see Fig.~\ref{fig:phase-diagram-BNNNI}F). 
\end{itemize} 
Ising-like transition points between the FM and homogeneous PM phases Fig.~\ref{fig:phase-diagram-BNNNI} were further determined from the 
asymptotic finite-size scaling of the magnetic susceptibility and second moment correlation length, using the known critical exponents, as detailed in App.~\ref{app:fss}. Precise determinations of the other phase transitions were not attempted.

In the following, we apply the WCRG to representative points spanning all five phases of the sBNNNI model and to the Ising-like critical line, with particular emphasis on the latter.

\section{Wavelet conditional renormalization group}
\label{sec:methods}

The WCRG algorithm is based on learning the conditional probability
distributions of the fast degrees of freedom, represented by wavelet
coefficients, conditioned on the slow degrees of freedom, represented by
coarse-grained fields at different scales. A microscopic field configuration is recursively decomposed by an orthogonal
wavelet transform. At scale \(j\), the field \(\phi_{j-1}\), defined on a
lattice of linear size \(L_{j-1}\), is mapped to a coarse field \(\phi_j\) on a
lattice of linear size \(L_j=L/2^j\), together with a set of wavelet variables
\(\bar\phi_j\) containing the fast fluctuations removed by the coarse graining.
The explicit construction of the two-dimensional transform
\(\mathcal W_j\phi_{j-1}=(\phi_j = W_j \phi_{j-1},\bar\phi_j = \overline{W}_j\phi_{j-1})\), including the low-pass map
\(W_j\), the high-pass maps \(\overline W_j\), and the three wavelet channels,
is given in App.~\ref{app:wavelet-conventions}, see Eqs.~\eqref{eq:app-2d-coarse}--\eqref{eq:app-2d-transform-compact}.

The transform is orthogonal and therefore invertible. We denote its inverse by
\begin{equation}
\phi_{j-1}
=
\mathcal W_j^{-1}(\phi_j,\bar\phi_j).
\label{eq:W-1-filter}
\end{equation}
As discussed in App.~\ref{app:wavelet-conventions}, orthogonality also implies
that the corresponding change of variables has unit Jacobian. As a result, expressing probability densities in terms of \((\phi_j,\bar\phi_j)\) does not require any additional Jacobian factor.

Iterating the decomposition gives a one-to-one representation of the
microscopic field in terms of the coarsest field and all wavelet variables,
\begin{equation}
\phi_0
\longleftrightarrow
\left(
\phi_J,
\bar\phi_J,
\bar\phi_{J-1},
\ldots,
\bar\phi_1
\right).
\label{eq:multiscale-representation}
\end{equation}
In the numerical applications considered in this work, the hierarchy is
iterated down to \(L_J=1\).

The probability distribution of the field at scale \(j\) is written in terms of
an effective energy \(E_j\), 
\begin{equation}
p_j(\phi_j)
=
\frac{1}{Z_j}
e^{-E_j(\phi_j)}.
\label{eq:coarse-probability}
\end{equation}
That energy is obtained recursively from the finer-scale distribution (starting from the microscopic energy \(E_0\))
by integrating out the wavelet variables,
\begin{equation}
e^{-E_j(\phi_j)}
=
\int {\rm d}\bar\phi_j\,
e^{-E_{j-1}(\mathcal W_j^{-1}(\phi_j,\bar\phi_j))}.
\label{eq:rg-recursion}
\end{equation}
Note that here and in the following, effective energies are defined up to additive
constants, which are absorbed into the corresponding normalizations.

Because the wavelet transform has unit Jacobian, the joint probability density of
\((\phi_j,\bar\phi_j)\) is simply
\[
p_{j-1}\!\left(
\mathcal W_j^{-1}(\phi_j,\bar\phi_j)
\right).
\]
Therefore, the conditional distribution of the wavelet variables at fixed
coarse field is
\begin{equation}
p_j(\bar\phi_j|\phi_j)
=
\frac{
p_{j-1}\!\left(
\mathcal W_j^{-1}(\phi_j,\bar\phi_j)
\right)
}{
p_j(\phi_j)
}.
\label{eq:conditional-ratio}
\end{equation}
Equivalently, this conditional probability can be written as
\begin{equation}
p_j(\bar\phi_j|\phi_j)
=
\exp\left[
-\bar E_j(\bar\phi_j;\phi_j)
+
\bar F_j(\phi_j)
\right], 
\label{eq:conditional-prob}
\end{equation}
where the conditional energy is defined from the finer-scale effective energy,
\begin{equation}
\bar E_j(\bar\phi_j;\phi_j)
=
E_{j-1}\!\left(
\mathcal W_j^{-1}(\phi_j,\bar\phi_j)
\right).
\label{eq:def-barE}
\end{equation}

The corresponding free-energy contribution is fixed by normalization,
\begin{equation}
e^{-\bar F_j(\phi_j)}
=
\int {\rm d}\bar\phi_j\,
e^{-\bar E_j(\bar\phi_j;\phi_j)}.
\label{eq:def-barF}
\end{equation}

The exact microscopic distribution can be factorized into a product of
conditional probabilities associated with the successive wavelet decompositions of 
Eq.~\eqref{eq:multiscale-representation},
\begin{equation}
p_0(\phi_0)
=
p_J(\phi_J)
\prod_{j=1}^{J}
p_j(\bar\phi_j|\phi_j).
\label{eq:exact-multiscale-factorization}
\end{equation}
Equivalently, using Eqs.~\eqref{eq:coarse-probability} and
\eqref{eq:conditional-prob}, the microscopic energy can be expressed using a
hierarchical form,
\begin{equation}
E_0(\phi_0)
=
E_J(\phi_J)
+
\sum_{j=1}^{J}
\left[
\bar E_j(\bar\phi_j;\phi_j)
-
\bar F_j(\phi_j)
\right],
\label{eq:exact-hierarchical-energy}
\end{equation}
neglecting an irrelevant additive constant.
Up to this point, everything is exact. The idea behind WCRG is that expressing the probability distributions as a cascade over conditional probability distributions, Eq.~\eqref{eq:multiscale-representation}, allows one to circumvent critical slowing down since all $p_j(\bar\phi_j|\phi_j)$s are easy to sample. 

The challenge is then to learn or infer from data these conditional distributions. In order to do that,
 the exact energies appearing in
Eq.~\eqref{eq:exact-hierarchical-energy} are estimated by finite-dimensional
parametrizations. The coarsest energy \(E_J(\phi_J)\) is approximated by a
parametric energy \(E_{\theta_J}(\phi_J)\), where \(\theta_J\) is the vector of
learned coefficients. Because we here set \(L_J=1\), \(E_{\theta_J}\) is a
one-dimensional potential expanded on a fixed finite family of sigmoid
functions, as described in App.~\ref{app:conditional-learning}.

Each exact conditional energy \(\bar E_j(\bar\phi_j;\phi_j)\) is approximated as
\begin{equation}
\bar E_{\bar\theta_j}(\bar\phi_j;\phi_j)
=
\bar\theta_j^{\mathrm T}
\Psi_j( W_j^{-1}(\phi_j,\bar\phi_j)),
\label{eq:conditional-energy-ansatz}
\end{equation}
where \(\Psi_j\) collects the retained symmetry-allowed functions of the
fields. In the present work,
\(\Psi_j\) contains translationally invariant quadratic couplings and local
potential terms, as detailed in App.~\ref{app:conditional-learning}. The vector
\(\bar\theta_j\) contains the associated coupling constants. The ansatz is
therefore linear in the parameters \(\bar\theta_j\), while the functions in
\(\Psi_j\) can be nonlinear functions of the fields. The parameters \(\bar\theta_j\) are learned independently at each scale by
matching the score to the conditional distributions
\(p_j(\bar\phi_j|\phi_j)\).

After learning, synthetic microscopic configurations are generated
recursively. One first samples the coarsest field from
\(p_{\theta_J}(\phi_J)\), and then for \(j=J,J-1,\ldots,1\) one samples
\(\bar\phi_j\) from the learned conditional distribution at fixed \(\phi_j\)
and reconstructs the finer field through Eq.~\eqref{eq:W-1-filter}. The
conditional sampling is performed using a standard local Metropolis Monte Carlo
algorithm. At fixed \(\phi_j\), the acceptance ratio for a proposed change
\(\bar\phi_j\to\bar\phi_j'\) depends only on
\begin{equation}\label{eq:delta-barE}
\Delta\bar E_{\bar\theta_j}
=
\bar E_{\bar\theta_j}(\bar\phi_j';\phi_j)
-
\bar E_{\bar\theta_j}(\bar\phi_j;\phi_j).
\end{equation}
Therefore, for the purpose of generating synthetic configurations,
only the coarsest energy \(E_{\theta_J}\) and the conditional energies
\(\bar E_{\bar\theta_j}\) are needed explicitly. (See App.~\ref{app:conditional-learning} for implementation details.)

The learned conditional energy
\(\bar E_{\bar\theta_j}\) nevertheless induces a conditional normalization,
\begin{equation}
\bar F_{\bar\theta_j}^{\rm ind}(\phi_j)
=
-\log
\int {\rm d}\bar\phi_j\,
\exp\left[
-\bar E_{\bar\theta_j}(\bar\phi_j;\phi_j)
\right].
\label{eq:learned-induced-free-energy}
\end{equation}
Although this quantity is not computed during conditional sampling, it defines the
synthetic hierarchical measure actually sampled by the generative procedure,
\begin{equation}
E_0^{\rm synth}(\phi_0)
=
E_{\theta_J}(\phi_J)
+
\sum_{j=1}^{J}
\left[
\bar E_{\bar\theta_j}(\bar\phi_j;\phi_j)
-
\bar F_{\bar\theta_j}^{\rm ind}(\phi_j)
\right].
\label{eq:synth-hierarchical-energy}
\end{equation}
In general, this energy differs from the original microscopic energy
\(E_0\), because the conditional energies are represented within a finite
ansatz class.

To reconstruct an explicit microscopic energy functional
associated with the learned WCRG hierarchy, the induced free-energy terms must
also be approximated. We denote these fitted free-energy contributions
\(\bar F_{\alpha_j}\), where \(\alpha_j\) denotes the corresponding vector of
parameters. This gives the explicit WCRG energy
\begin{equation}
E_0^{\rm WCRG}(\phi_0)
=
E_{\theta_J}(\phi_J)
+
\sum_{j=1}^{J}
\left[
\bar E_{\bar\theta_j}(\bar\phi_j;\phi_j)
-
\bar F_{\alpha_j}(\phi_j)
\right].
\label{eq:wcrg-hierarchical-energy}
\end{equation}
(See App.~\ref{app:conditional-learning} for details of the \(\alpha_j\) estimation.)

The choice of wavelet family is adapted to the phase structure. We use Haar wavelets \cite{Haar1910} for the points that have an homogeneous texture, namely within the
FM and PM phases, and along Ising-like critical line. 
We use Daubechies wavelets with four
vanishing moments (DB4) \cite{Daubechles+2009+564+652} for the MPM, IC, and AF regimes. The smoother and wider filters of the latter are better suited to finite-wavevector modulations.

\section{Numerical results}\label{sec:numerical-results}
In this section, we test the efficacy of the WCRG construction as a sampling
method for the sBNNNI model. We first compare Monte Carlo and WCRG-generated
configurations across representative regions of the phase diagram. We then
analyze the decorrelation of the conditional wavelet chains at criticality,
where local Monte Carlo dynamics suffers from critical slowing down. Finally,
we study configuration-wide observables, focusing on the magnetization and the
microscopic energy as diagnostics of the generated measure.

\subsection{Wavelet sampling across the phase diagram}\label{sec:wavelet-sampling}
\begin{figure*}
    \centering
    \includegraphics[width=0.99\linewidth]{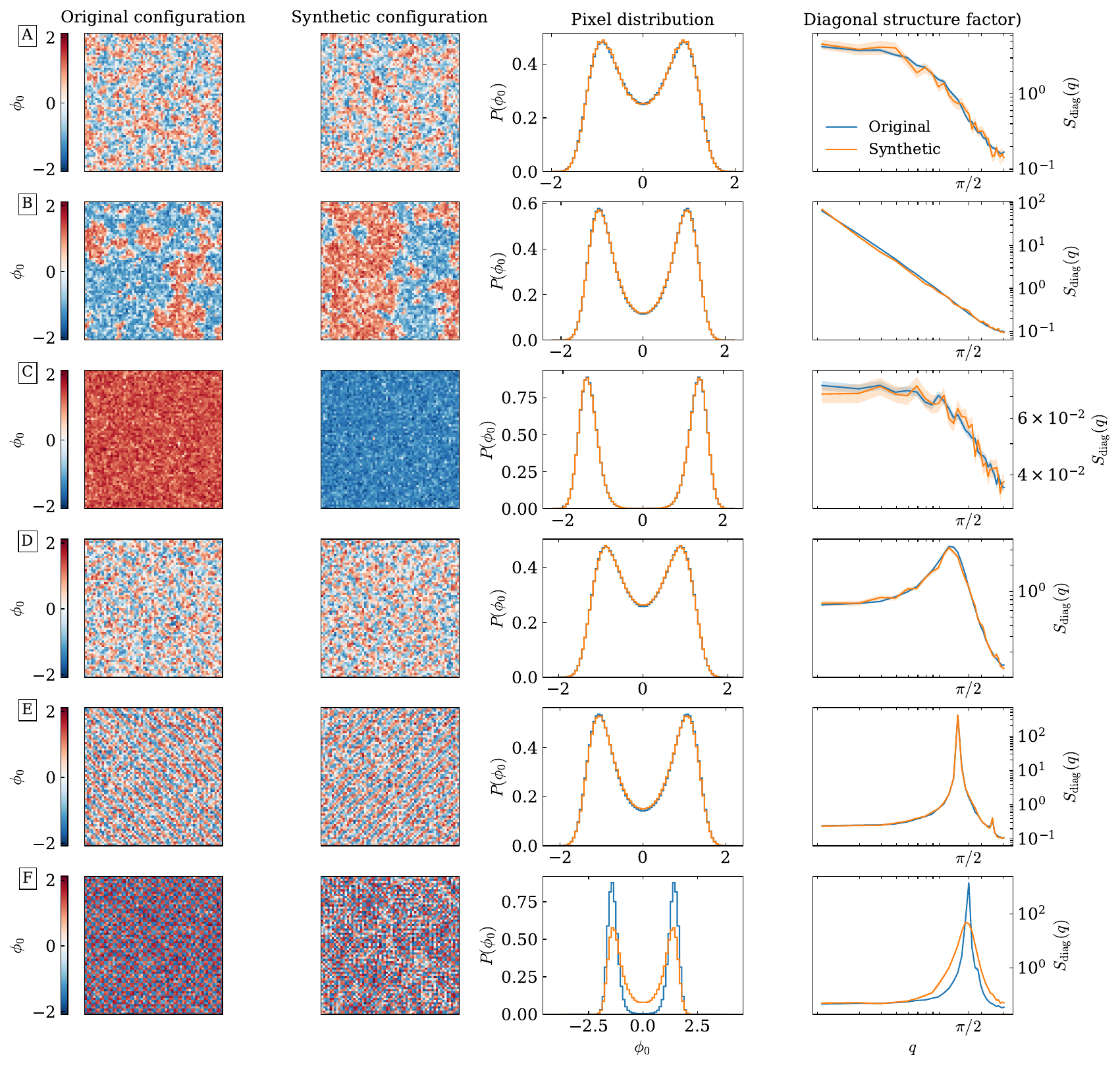}
    \caption{\small 
Comparison between Monte Carlo configurations and samples generated through WCRG for representative points of the sBNNNI phase diagram (A--F in Fig.~\ref{fig:phase-diagram-BNNNI}). 
Subsequent rows correspond to a point---from top to bottom: (A) in the PM phase, (B) along the Ising critical line, (C) in the FM phase, (D) in the MPM phase, (E) in the IC phase, and (F) in the AF phase. 
For each point, the panels display: (1st column) an original Monte Carlo configuration, 
(2nd column) a synthetic configuration 
reconstructed through recursive conditional sampling,
the comparison between the original (blue) and the reconstructed (yellow) distributions of (3rd column) \(P(\phi_0(i))\) and (4th column) the diagonal structure factor \(S_{\rm diag}(q)\) (see Eq.~\eqref{eq:Sdiag}). 
Standard Haar wavelets are used for the FM, Ising critical, and PM points. To better capture the emerging modulated structures, Daubechies wavelets with four vanishing moments (DB4) are used for the MPM, IC, and AF phases. 
A good agreement is obtained between generated and original samples for both local observables and long-range correlations in all cases except the AF phase, for which the reconstruction is hindered for the reasons explained at the end of Sec.~\ref{sec:wavelet-sampling}.
}
    \label{fig:numerical-results}
\end{figure*}
We apply the WCRG construction to the representative points of the sBNNNI
phase diagram in Fig.~\ref{fig:phase-diagram-BNNNI}. Unless otherwise stated,
we use systems of linear size \(L=64\). For each condition \((T,\kappa)\), we
first generate a reference ensemble of 3000 equilibrium Monte Carlo configurations
of the microscopic model in Eq.~\eqref{eq:Hsoft}, using local single-site
Metropolis updates (see App.~\ref{app:conditional-learning}). These configurations are then used both as reference samples and as training data for the conditional WCRG models.

For the ordered phases, namely FM, IC, and AF, each run is initialized from an
independent random configuration. The system is then equilibrated until the
magnetization \(m_0\) and the second moment \(\langle\phi_0^2\rangle\), defined
in Eqs.~\eqref{eq:def-m} and \eqref{eq:def-phiq}, reach stationary values, and
discard an additional time window of at least five autocorrelation times of
these observables. Simulation time is defined in units of Monte Carlo sweeps, which consist of \(L^2\) attempted single-site updates. In the PM and MPM phases, and at the Ising-like critical
point, we start from a disordered configuration and sample configurations only
after the same observables have thermalized. Configurations separated in time by at least twice the largest measured autocorrelation time among these observables are then stored for further analysis. At the critical point, this procedure reflects the critical slowing down of the underlying local Metropolis dynamics: for \(L=64\), the largest autocorrelation time is approximately \(10^4\) Monte Carlo sweeps, as shown in Fig.~\ref{fig:tau-criticality}. Hence, generating the 3000 configurations used here required up to  \(6\times10^7\) Monte Carlo sweeps, following equilibration. 

These Monte Carlo configurations are then used to construct the scale-dependent training pairs of Eq.~\eqref{eq:training-pairs}, from which the conditional WCRG models are learned. After training, (synthetic) configurations are generated---recursively from coarse to fine scales---according to the procedure described in Sec.~\ref{sec:methods}. 

Figure~\ref{fig:numerical-results} compares original and synthetic
configurations at representative points for each phase, which includes a typical microscopic configuration, the distribution of local field values
\(P(\phi_0(i))\), and the diagonal structure factor \(S_{\rm diag}(q)\) of Eq.~\eqref{eq:Sdiag}. The
local field distribution assesses whether the reconstruction captures the
single-site statistics of the soft-spin variable, while the structure factor
probes spatial correlations and, in particular, the dominant ordering
wavevector. 

Agreement is generally good for all phase points considered. In the homogeneous
phases and at the critical point, the reconstructed samples reproduce both the
local field statistics and the peak structure at small wavevector. In the MPM
and IC regimes, the method similarly captures the finite-wavevector peak of
\(S_{\rm diag}(q)\), showing that the synthetic configurations retain the
modulated structure of the original ones.

The AF regime exhibits the most pronounced discrepancy. This case offers a particularly stringent
test for the present reconstruction scheme, because the ordered state is a
rigid commensurate modulation whose wavelength is tied directly to the
microscopic lattice scale. In this situation, the low-pass fields generated at
intermediate scales carry little information about the alternating pattern. After coarse graining, the checkerboard modulation is largely averaged out;
the relevant information is concentrated only in the finest wavelet degrees of
freedom. As a result, the conditional reconstruction 
exploits the multiscale hierarchy much less effectively than in phases where the
dominant structures are visible at coarse and intermediate scales as well.

\subsection{Absence of critical slowing down}\label{sec:no-csd}

\begin{figure*}
    \centering
    \includegraphics[width=0.8\linewidth]{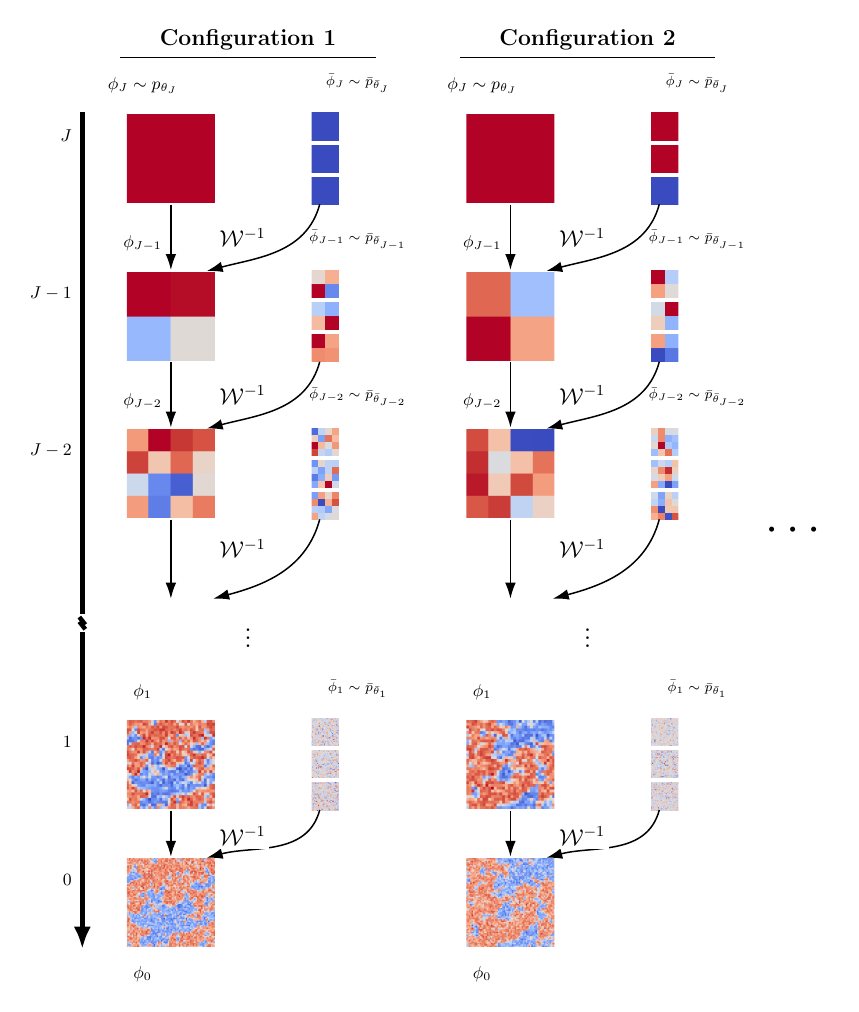}
    \caption{\small 
    Multiscale sampling and reconstruction of two representative
    configurations. Each realization is generated starting from the coarsest scale (top configurations, $\phi_J$) and moving towards the finer ones (bottom configurations, $\phi_0$) by accounting recursively for the details encoded in three sampled wavelets at each scale (denoted $\bar \phi_J, \ldots,\bar\phi_1$ and represented by smaller squares). Colors encode $\phi_j$ and $\bar\phi_j$ at the various levels $j=J, J-1, \ldots, 0$, as in the previous figures. 
    In practice, the coarsest variable \(\phi_J\) is sampled first. Then, at each scale
    \(j=J,J-1,\ldots,1\), the wavelet variables \(\bar\phi_j\) are sampled
    from the learned conditional distribution at fixed \(\phi_j\), and the
    next finer field \(\phi_{j-1}\) is reconstructed through the inverse
    wavelet transform ${\cal W}^{-1}$. The large-scale structure of the eventual finer configuration $\phi_0$ is therefore already determined at
    the coarser levels and is progressively dressed by shorter-wavelength
    fluctuations, rather than assembled through local propagation on the
    microscopic lattice. Note that at the coarsest scale configurations 1 and 2 are rather similar and so are two of the three initial wavelets $\bar \phi_J$. Minor differences on that scale nevertheless suffice to give rise to completely different clusters at the finest scale.}
    \label{fig:sampling-scheme}
\end{figure*}

At a continuous phase transition, critical slowing down originates from the
growth of the correlation length \(\xi\). In a finite system at criticality,
\(\xi\) becomes comparable to the system size \(L\). For MCMC algorithms based on local microscopic updates, the reorganization of
correlated structures then requires information to propagate across distances
of order \(L\), leading to a growth of equilibration and decorrelation times
with system size \cite{RevModPhys.49.435,Sokal1997}.

To quantify this effect, we consider an observable \(O(t)\) measured along a
trajectory, with time \(t\) measured in Monte Carlo sweeps, 
for a step size chosen such that the acceptance rate is about $50\%$.
The normalized autocorrelation function in the stationary state is
\begin{equation}
C_O(t)
=
\frac{
\langle O(t_0+t)O(t_0)\rangle
-
\langle O\rangle^2
}{
\langle O^2\rangle
-
\langle O\rangle^2
},
\label{eq:autocorrelation-function}
\end{equation}
where the average is over the reference time \(t_0\) along the stationary
trajectory. At long times we fit
\begin{equation}
C_O(t)
\sim
e^{-t/\tau_O},
\label{eq:exponential-autocorrelation}
\end{equation}
which defines the exponential autocorrelation time \(\tau_O\). This quantity
provides an estimate of the number of sweeps needed to obtain effectively
decorrelated samples with respect to the observable \(O\).

\begin{figure}[t]
    \centering
    \includegraphics[width=0.93\columnwidth]{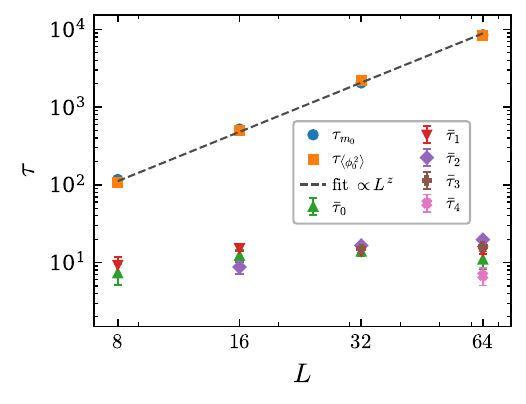}
    \caption{\small
    Autocorrelation times at the Ising-like critical point of the sBNNNI model
    with \(\kappa=0.2\) and \(T=T_{\rm pc}(L)\). For local Monte Carlo dynamics, the autocorrelation times of the
    magnetization \(m_0\) and of the second moment
    \(\langle\phi_0^2\rangle\) grow algebraically with system size. The
    dashed line shows a power-law fit, yielding
    \(z_{\rm loc}=2.11(4)\). By contrast, the conditional autocorrelation
    times \(\bar\tau_j(L)\), measured for the second moment of the
    reconstructed field during conditional wavelet sampling at scale \(j\),
    remain of order unity. No systematic growth is observed either with the
    final microscopic size \(L\) or with the scale-dependent size
    \(L_j=L/2^j\).
    }
    \label{fig:tau-criticality}
\end{figure}

For conventional local MCMC dynamics at criticality, autocorrelation
times are expected to grow algebraically,
\begin{equation}
\tau_O
\sim
L^{z_O},
\label{eq:tau-algebraic-csd}
\end{equation}
where the dynamical exponent may depend on the update rule and on the observable.
As a microscopic reference dynamics, we use single-site local Monte Carlo
updates for the sBNNNI model at \(\kappa_x=\kappa_y=0.2\) and at the
size-dependent (pseudo-)critical temperature \(T_{\rm pc}(L)\). 
In order to investigate the critical slowing down,
we naturally monitor both the magnetization
\begin{equation}
m_0
=
\frac{1}{L^2}
\sum_i
\phi_0(i),
\label{eq:def-m}
\end{equation}
and the spatial second moment of the field,
\begin{equation}
\langle \phi_0^2\rangle
=
\frac{1}{L^2}
\sum_i
\phi_0^2(i).
\label{eq:def-phiq}
\end{equation}
The latter is here particularly important because magnetization---the standard observable in these systems---is ill-suited to extract an autocorrelation time in dynamical algorithms based on WCRG, for reasons discussed below. As shown in Fig.~\ref{fig:tau-criticality}, both observables display the
expected algebraic growth. A power-law fit gives
\begin{equation}
\tau_{m_0},
\,
\tau_{\langle\phi_0^2\rangle}
\sim
L^{z_{\rm loc}},
\qquad
z_{\rm loc}=2.11(4),
\label{eq:tau-local-critical}
\end{equation}
compatible, within the accuracy of our simulations, with the known dynamic
critical exponent of local single-spin dynamics in the two-dimensional Ising
universality class, $z = 2.1665(12)$~\cite{PhysRevLett.76.4548, PhysRevB.62.1089}.

Cluster algorithms provide an important point of comparison. In unfrustrated
ferromagnetic systems, Swendsen--Wang and Wolff updates construct extended
correlated clusters and update them collectively, thereby strongly reducing the
critical slowing down associated with single-site dynamics
\cite{Sokal1997,PhysRevLett.61.2635}. In frustrated systems, however,
constructing such clusters is not directly possible~\cite{Zheng2022,Alfaro2026}. The WCRG strategy
is conceptually different. It does not attempt to identify microscopic clusters
to be flipped, but instead samples the long-wavelength degrees of freedom first
and then conditionally generates the fluctuations associated with progressively
finer wavelet degrees of freedom. The sampling protocol is described in Sec.~\ref{sec:methods} and illustrated in Fig.~\ref{fig:sampling-scheme}, which shows the scale-by-scale reconstruction of two representative configurations. As the figure illustrates, the main spatial structures present in the final fields \(\phi_0\) can already be recognized at intermediate scales, while each successive reconstruction step adds ever finer details. The two configurations are not obtained by slowly decorrelating one microscopic field from the other, but through two separate top-down cascades, with independently sampled wavelet variables introducing different fluctuations at every level. The statistical efficiency of this construction is quantified below through the autocorrelation times of the conditional wavelet chains.

In this context, magnetization is not a useful observable for monitoring the evolution of the conditional chain. For the wavelet transform used here, each wavelet channel contains at least one high-pass filter whose coefficients sum to zero, as expressed by Eq.~\eqref{eq:app-filter-sums} in App.~\ref{app:wavelet-conventions}. Consequently, the contribution of the wavelet variables \(\bar\phi_j\) to the reconstructed finer field has zero spatial average. Since the conditional dynamics updates only \(\bar\phi_j\) while keeping \(\phi_j\) fixed, the spatial average of the reconstructed field is entirely determined by \(\phi_j\), up to the fixed normalization factor, and remains constant along the conditional chain. The magnetization therefore carries no information about the decorrelation of this chain. We instead monitor the spatial second moment of the reconstructed field,
\begin{equation}
O_{j,L}(t)
=
\frac{1}{L_{j-1}^{2}}
\sum_{i\in\Lambda_{j-1}}
\phi_{j-1}^{2}(i,t),
\label{eq:conditional-second-moment}
\end{equation}
where \(L\equiv L_0\) is the final microscopic size and
\(L_j=L/2^j\) is the size of the coarse field at reconstruction level \(j\).
For each pair \((L,j)\), we measure the autocorrelation function of
\(O_{j,L}(t)\) along the conditional Markov chain and extract the corresponding
exponential autocorrelation time, \(\bar\tau_j(L)\).  The reported values are obtained by fitting the autocorrelation function averaged over 50 independent chains, with error bars indicating the corresponding statistical uncertainty.

The results in Fig.~\ref{fig:tau-criticality} show a qualitative difference
between local and WCRG dynamics, which is one of the main results of this work. Within the
range of sizes and scales investigated here, the conditional autocorrelation
times remain bounded,
\begin{equation}
\bar\tau_j(L)
=
\mathcal O(1),
\label{eq:conditional-autocorrelation-o1}
\end{equation}
with no detectable systematic increase as either \(L\) or \(L_j\) grows.
In other words, although the conditional distributions are sampled with
local Metropolis moves, a fixed number of conditional sweeps suffices to
decorrelate wavelet variables at each reconstruction level.
A complete microscopic realization the requires but one conditional sampling stage for
each wavelet level. Iterating the hierarchy down to \(L_J=1\), requires sampling all levels,
\begin{equation}
J
=
\log_2 L.
\label{eq:number-wavelet-levels}
\end{equation}
As a result, the sequential decorrelation depth of the full WCRG reconstruction,
measured in conditional-sweep units, scales as
\begin{equation}
\tau_{\rm WCRG}^{\rm depth}
\sim
\sum_{j=1}^{J}
\bar\tau_j(L)
=
\mathcal O(J)
=
\mathcal O(\log_2 L).
\label{eq:tau-wcrg-depth}
\end{equation}
This logarithmic scaling should not be confused with the total number of
elementary local update attempts. A conditional sweep at scale \(j\) updates a
number of wavelet degrees of freedom proportional to \(L_j^2\). If
\(\bar\tau_j(L)=\mathcal O(1)\), the total number of local update attempts required for the full hierarchy scales as
\begin{equation}
\mathcal \#_{\rm WCRG}
\propto
\sum_{j=1}^{J}
L_j^2
=
\mathcal O(L^2),
\label{eq:wcrg-elementary-cost}
\end{equation}
up to scale-independent factors. This cost is
\emph{linear} in the number of microscopic degrees of freedom and does not acquire the
\emph{additional} critical scaling factor \(L^{z_O}\)  characteristic of local microscopic
sampling.

%
%
%

\subsection{Configuration-wide observables}
\label{sec:config-wise-obs}

In the two previous subsections, 
we demonstrated the ability of the WCRG to generate microscopic configurations across different phases of the sBNNNI model and along its Ising-like critical line. The quality of the generated samples was assessed by comparing local observables, i.e., the single-site probability distribution $P(\phi_0(i))$ and the diagonal structure factor $S_{\rm diag}(q)$ which probes two-point correlation functions.

\begin{figure}[t]
\centering
    \begin{overpic}[width=\linewidth]{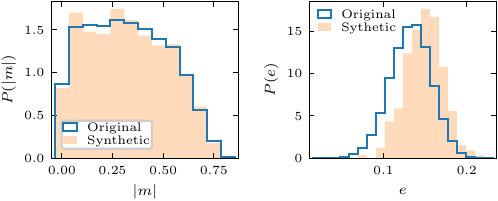}
  \put(5,42){(a)}
  \put(55,42){(b)}
\end{overpic}
    \caption{\small 
Comparison between the statistics of configuration-wide 
observables for original and synthetic configurations 
along the Ising-like critical point of the sBNNNI model, with \(\kappa=0.2\), \(T=1.0\), and \(L=64\). 
Distribution 
of (a) the absolute magnetization \(|m|\) and (b) the intensive microscopic energy $e$. While the former  shows excellent
agreement between original and synthetic samples
the latter presents a
visible mismatch. 
This discrepancy reflects the truncation of the exact multiscale effective energies entering the hierarchical  probability distribution of the WCRG. %
}
    \label{fig:magn-ene-hist}
\end{figure}

We now consider configuration-wide observables. The simplest such examples are the magnetization $m$ from Eq.~\eqref{eq:def-m}, and the energy per site
\begin{equation}
e
=
\frac{E_0[\phi_0]}{L^2}.
\label{eq:def-e}
\end{equation}
Figure~\ref{fig:magn-ene-hist} compares these two quantities for original and synthetic configurations along the Ising-like critical point (with $\kappa=0.2$, $T=1.0$, and $L=64$). 

\paragraph*{Magnetization.}
Figure~\ref{fig:magn-ene-hist}(a) shows excellent agreement between the magnetization distributions obtained on these two different kinds of configurations.
%
%
However, this agreement should be interpreted with caution.
The distribution \(P(m)\) encodes all moments \(\langle m^n\rangle\), which are actually determined by spatial integrals of the multi-point correlation functions of the microscopic variables.
Accordingly, once the complete hierarchy of correlation functions is reproduced, the magnetization distribution $P(m)$ follows automatically. The converse, however, is not true: reproducing the magnetization distribution $P(m)$ alone does not guarantee the correct reproduction of the full hierarchy of correlation functions.

Reproducing the magnetization distribution 
in Fig.~\ref{fig:magn-ene-hist}(a) is therefore largely built into the multiscale reconstruction scheme of the WCRG. 
At the coarsest scale \(j=J\), the field \(\phi_J\) consists of a single degree of freedom \((L_J=1)\), which is proportional to the spatial average of the microscopic field (i.e., of the field $\phi_0$ at scale $j=0$). Learning its distribution amounts to fitting the one-dimensional effective potential $V_J(\phi_J)$ with \(E_J(\phi_J)=V_J(\phi_J)\). Once this coarse distribution is correctly reproduced, the subsequent recursive reconstruction preserves the total magnetization, up to the normalization convention of the wavelet transform, because all wavelet modes are constructed to have  vanishing spatial averages. As a result, the magnetization distribution is fixed at the coarsest scale and is preserved throughout the reconstruction.

\paragraph*{Energy.} 
Figure~\ref{fig:magn-ene-hist}(b) presents a visible mismatch between the two energy distributions. 
This discrepancy is rooted in the synthetic configurations not being drawn from the exact microscopic Gibbs measure, \(p_0(\phi_0)\). As discussed
in Sec.~\ref{sec:methods}, the exact hierarchical representation of the
microscopic energy, Eq.~\eqref{eq:exact-hierarchical-energy}, involves the true
conditional energies \(\bar E_j\), defined in Eq.~\eqref{eq:def-barE}, and the
corresponding free-energy terms \(\bar F_j\), defined in
Eq.~\eqref{eq:def-barF}. These quantities are determined by the exact
coarse-grained effective energies, which contain all the possible interaction terms compatible with the symmetry of the model, that are
generated by integrating out finer degrees of freedom in
Eq.~\eqref{eq:rg-recursion}.

By contrast, in the learned model the exact conditional energies \(\bar E_j\) are
estimated by the finite-dimensional ansatz
\(\bar E_{\bar\theta_j}\) of Eq.~\eqref{eq:conditional-energy-ansatz}. The
conditional sampling procedure therefore uses
\begin{equation}
\bar p_{\bar\theta_j}(\bar\phi_j|\phi_j)
=
\exp\left[
-\bar E_{\bar\theta_j}(\bar\phi_j;\phi_j)
+
\bar F_{\bar\theta_j}^{\rm ind}(\phi_j)
\right],
\label{eq:learned-conditional-energy-distribution}
\end{equation}
where the normalization \(\bar F_{\bar\theta_j}^{\rm ind}\) is the induced
free energy defined in Eq.~\eqref{eq:learned-induced-free-energy}. Because
$\bar E_{\bar\theta_j}$ is in general only an approximation of $\bar E_j$,
\(\bar F_{\bar\theta_j}^{\rm ind}\) is also different from the exact
\(\bar F_j\): the two quantities normalize different conditional energies.
Consequently,
\begin{equation}
\bar p_{\bar\theta_j}(\bar\phi_j|\phi_j)
\neq
\bar p_j(\bar\phi_j|\phi_j),
\end{equation}
even for an infinite number of training samples. In short, our chosen finite-dimensional representation of $\bar E_j$ is not expressive enough.
The reconstruction procedure therefore samples from the synthetic hierarchical
measure defined in Eq.~\eqref{eq:synth-hierarchical-energy}, rather than from
the exact Gibbs measure associated with the microscopic energy \(E_0\). In
general,
\begin{equation}
p_0^{\rm synth}(\phi_0)\neq p_0(\phi_0),
\qquad
E_0^{\rm synth}(\phi_0)\neq E_0(\phi_0).
\end{equation}

In other words, the
distribution is obtained by evaluating the original microscopic energy \(E_0\) on
configurations generated from the synthetic measure. Because these configurations
are not distributed exactly according to \(p_0\), the distribution of
\(E_0[\phi_0^{\rm synth}]\) need not coincide with the distribution of
\(E_0[\phi_0^{\rm orig}]\). 
Resolving the mismatch in Fig.~\ref{fig:magn-ene-hist}(b) would require a more refined finite-dimensional representation of $\bar E_j$, but such representation is not immediately available. This issue is therefore left as future work.

\section{Conclusion and 
perspectives}
\label{sec:conclusions}

In this work we applied the WCRG to the
sBNNNI model as a benchmark for learned multiscale sampling in frustrated
systems. As described in Sec.~\ref{sec:methods}, the method decomposes
microscopic configurations into coarse fields and wavelet degrees of freedom,
learns the conditional distribution of the wavelet variables at each scale, and
generates new configurations recursively from coarse to fine scales.

In Sec.~\ref{sec:wavelet-sampling}, we showed that WCRG-generated
configurations reproduce the main statistical features of Monte Carlo samples
across representative points of the sBNNNI phase diagram. In particular, the
method captures both local field distributions and diagonal structure factors
in the homogeneous phases, along the Ising-like critical line, and in the
modulated MPM and IC regimes. The AF regime, however, is pathological for this approach. 

In Sec.~\ref{sec:no-csd}, we showed that, while conventional local Monte Carlo
dynamics displays the expected critical slowing down, the conditional wavelet
chains used in WCRG remain decorrelated in \(\mathcal O(1)\) sweeps at each
reconstruction scale, within the range of sizes investigated here. Because the
number of wavelet levels grows as \(\log_2 L\), the computational time required to generate independent configurations grows only logarithmically with \(L\). The total
number of elementary local updates remains proportional to the number of
generated degrees of freedom, but does not require an additional critical
factor \(L^z\) characteristic of local microscopic sampling. For the sBNNNI model, this represents a significant
sampling advantage over standard Monte Carlo strategies: local updates suffer
from critical slowing down, while cluster algorithms, which are highly
efficient in unfrustrated systems, cannot be  exploited in the
presence of frustration, even in the weakly frustrated regime.

In Sec.~\ref{sec:config-wise-obs}, we analyzed configuration-wide observables.
Although the magnetization distribution is accurately reproduced ---by construction---the microscopic
energy histogram presents a residual mismatch. This difference reflects the main
tradeoff of the method: Monte Carlo sampling is asymptotically exact but slows
down near criticality, whereas WCRG provides fast sampling but from an approximate
recursive measure whose accuracy is controlled by the expressiveness of the
parametrized energies.

From this perspective, WCRG is best viewed as a complementary strategy to
Monte Carlo. It trades exact sampling at fixed microscopic energy for a fast
generative procedure that can, in principle, be systematically improved. The
present results suggest that this tradeoff can be particularly useful in
frustrated systems, where conventional cluster constructions are ineffective.

This places WCRG within the broader line of machine-learning-assisted sampling
approaches in statistical physics
\cite{Gabrie2022,Wu2019,Ghio2024,DelBono2025,DelBono2026}. Its specific
contribution is to combine learning with an explicit multiscale RG structure:
the learned functions are  the conditional energies associated with each length
scale, rather than a single black-box generator. This organization makes it
possible to associate approximation errors with specific reconstruction levels
and to improve the model in a controlled way.

Natural directions for further studies include adding symmetry-allowed operators beyond the
quadratic-plus-local ansatz used in the present work. 
Developing better ansätze would also be expected to resolve the poor sampling of AF phase.
Another possibility,
which provides a compromise between expressiveness and interpretability, is to
consider convolutional neural networks with local receptive fields, following
Ref.~\cite{kadkhodaie2023learning}. Quantifying how WCRG errors decrease with
ansatz expressiveness, and developing practical diagnostics for this
convergence, are important directions for future work.

Finally, the scaling analysis in Sec.~\ref{sec:no-csd} assumes that the conditional energies were already learned from an available set original configurations. To achieve a given accuracy in this learning, one might need to analyze an increasing number of training configurations as their size increases, possibly leading to an additional increase of the computational cost with $L$.   We expect this dependence to be rather weak, but a systematic evaluation of this effect for WCRG, remains to be considered and is left for future investigation.

\paragraph*{Data availability statement.}
The data that support the findings of this article are openly available~\cite{RDR}. [Data will be made openly available upon manuscript acceptance, but a DOI is not available at the time of submission.]

\section*{Acknowledgment}

We thank Etienne Lempereur for helpful explanations of the implementation of
the WCRG algorithm and of the code associated with
Ref.~\cite{pmlr-v202-guth23a} that we used for this work. Gabriele Bandini thanks Cristiano Muzzi for helpful discussions throughout the development of this work.

\appendix

\section{Low-temperature analysis of the soft-spin BNNNI model}
\label{app:T0_BNNNI}

In this section, we analyze the low-temperature behavior of BNNNI model. For this purpose, we distinguish between the discrete-spin---\(\lambda\to\infty\)---BNNNI model (dBNNNI), whose fields
take the values \(\phi_i=\pm1\), and the soft-spin---finite $\lambda$---sBNNNI model considered in
this work, for which \(\phi_i\in\mathbb R\). 
Note, however, that the relation
between the hard-spin limit and the low-temperature analysis below requires
some care, because the latter is performed at fixed finite \(\lambda\).

An important qualitative difference between the phase diagram in
Fig.~\ref{fig:phase-diagram-BNNNI} and that of the discrete model studied in
Ref.~\cite{Hu2021_2} is the extent of the incommensurate (IC) phase. In the
soft-spin model, the IC regime is broad and appears to extend down to
relatively low temperatures, whereas in the discrete model it disappears
as \(T\to0\), where the ferromagnetic phase gives way directly to the
antiphase. Note that the purpose of the analysis below is not to determine the
two-dimensional phase boundaries quantitatively, but to identify a
low-temperature mechanism by which continuous field amplitudes at finite
\(\lambda\) can stabilize modulated profiles over an interval of frustration
strengths.

We consider the auxiliary one-dimensional analogue with dimensionless energy,
consistently with Eq.~\eqref{eq:Hsoft},
\begin{equation}
\begin{split}
E_T[\phi]
={}&
-\frac{J}{T}\sum_i\phi_i\phi_{i+1}
+\frac{\kappa J}{T}\sum_i\phi_i\phi_{i+2}
\\
&+\sum_i\left[
\phi_i^2+\lambda\left(\phi_i^2-1\right)^2
\right].
\end{split}
\label{eq:frust-app-1d-model}
\end{equation}
Because \(1/T\) multiplies only the quadratic interaction terms, the
low-temperature behavior is controlled by a balance between the quadratic
interaction energy and the quartic confinement. At the level of scaling,
\begin{equation}
\frac{\phi^2}{T}\sim\lambda\phi^4,
\end{equation}
so that
\begin{equation}
\phi^2\sim\frac{1}{\lambda T}.
\end{equation}
This scaling suggests introducing directly the rescaled field
\begin{equation}
\psi_i=\sqrt{\lambda T}\,\phi_i,
\label{eq:frust-app-rescaled-field}
\end{equation}
for expressing Eq.~\eqref{eq:frust-app-1d-model},
\begin{equation}
\begin{split}
E_T[\psi]
={}&\frac{1}{\lambda T^2}
\left[
-J\sum_i\psi_i\psi_{i+1}
+\kappa J\sum_i\psi_i\psi_{i+2}
+\sum_i\psi_i^4
\right]
\\
&+\frac{1-2\lambda}{\lambda T}\sum_i\psi_i^2
+\lambda N.
\end{split}
\label{eq:frust-app-low-temperature-expansion}
\end{equation}
For fixed finite \(\lambda\), the term in square brackets dominates as
\(T\to0\). Its prefactor is positive and therefore does not affect the
profile selected by minimization. The leading low-temperature variational
problem therefore reduces to minimizing the reduced functional
\begin{equation}
\mathcal F[\psi]
=-J\sum_i\psi_i\psi_{i+1}
+\kappa J\sum_i\psi_i\psi_{i+2}
+\sum_i\psi_i^4,
\label{eq:frust-app-leading-functional}
\end{equation}
which depends neither on \(T\) nor on \(\lambda\).
These two parameters determine only the relation between the original and
rescaled amplitudes, \(\phi_i=\psi_i/\sqrt{\lambda T}\), and the overall scale
of the leading energy. In particular, at fixed finite \(\lambda\), the
original field diverges as \(T^{-1/2}\), whereas the rescaled field remains
finite.

\subsection{Variational families and the leading \texorpdfstring{$(T\to0)$}{(T->0)} window}
\label{app:frust-leading}
We compare three variational families: the FM state is uniform,
\begin{equation}
\psi_i=m;
\end{equation}
the one-dimensional antiphase is
\begin{equation}
\psi_i=A(-1)^{\lfloor i/2\rfloor},
\label{eq:frust-app-antiphase-ansatz}
\end{equation}
corresponding to the repeating pattern \(++--\);
and the multiharmonic IC profile has
\begin{equation}
\psi_i
=A\sum_{n=0}^{N_h-1}c_{2n+1}
\cos\!\left[(2n+1)q_\star(\kappa)i\right],
\qquad c_1=1,
\label{eq:frust-app-multiharmonic-ansatz}
\end{equation}
with \(N_h=5\). The wavevector is chosen to minimize the quadratic part of
Eq.~\eqref{eq:frust-app-1d-model},
\begin{equation}
q_\star(\kappa)
=\arccos\!\left(\frac{1}{4\kappa}\right),
\qquad \kappa\geq\frac14.
\label{eq:frust-app-optimal-wavevector}
\end{equation}
The overall amplitude and the coefficients \(c_3,c_5,c_7,c_9\) are optimized
variationally. Fixing \(q=q_\star\) is an approximation, because nonlinear
terms can shift the optimal modulation wavevector.

For a generic profile \(\psi_i=A f_i\), the reduced functional per site is
\begin{equation}
\frac{\mathcal F}{N}=a_2[f]A^2+a_4[f]A^4,
\label{eq:functional}
\end{equation}
with
\begin{align}
a_2[f]
&=-J\langle f_i f_{i+1}\rangle
+\kappa J\langle f_i f_{i+2}\rangle,
\\
a_4[f]
&=\langle f_i^4\rangle.
\end{align}
When \(a_2<0\), minimization over the amplitude gives
\begin{equation}
A^2=-\frac{a_2}{2a_4},
\qquad
\frac{\mathcal F^{\rm min}}{N}
=-\frac{a_2^2}{4a_4},
\label{eq:frust-app-amplitude-minimization}
\end{equation}
while \(A=0\) otherwise. The FM and antiphase branches are therefore
\begin{align}
\frac{\mathcal F^{\rm FM}}{N}
&=-\frac{J^2(1-\kappa)^2}{4},
&&\kappa<1,
\\
\frac{\mathcal F^{\rm AF}}{N}
&=-\frac{J^2\kappa^2}{4}.
\end{align}

\begin{figure}[t]
    \centering
    \includegraphics[width=0.97\linewidth]{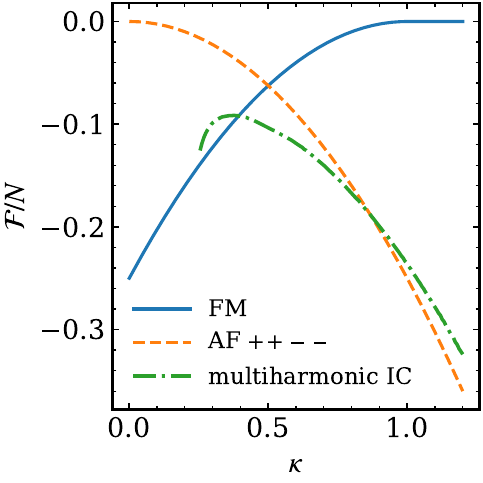}
    \caption{\small
    Reduced functional density \(\mathcal F/N\) of the FM state (solid line),
    period-four antiphase (dashed line), and multiharmonic IC profile
    (dash-dotted line). The IC branch is shown for \(\kappa\geq1/4\), where
    \(q_\star\) in
    Eq.~\eqref{eq:frust-app-optimal-wavevector} is real. Within the variational
    families considered, it has the lowest leading energy approximately for
    \(0.393\lesssim\kappa\lesssim0.869\). Since \(\mathcal F\) is independent
    of \(T\) and \(\lambda\), the profiles and crossing points are universal
    within the leading low-temperature scaling problem.}
    \label{fig:frust-app-zero-temperature}
\end{figure}

Figure~\ref{fig:frust-app-zero-temperature} compares these branches with the
optimized multiharmonic one. The IC
profile then has the lowest leading energy approximately for
\begin{equation}
0.393\lesssim\kappa\lesssim0.869.
\label{eq:frust-app-leading-IC-window}
\end{equation}
Because
\(\mathcal F\) contains neither \(T\) nor \(\lambda\), the three variational
profiles and their crossing points are independent of both parameters within
this leading scaling problem. The \(\lambda\) dependence enters only through
the common prefactor \(1/\lambda\) of the leading contribution to the original
energy in Eq.~\eqref{eq:frust-app-low-temperature-expansion}.
Note, however, that because this analysis only consider a finite set of variational families, this statement concerns the restricted ansatz space and does not establish the
global minimizer of Eq.~\eqref{eq:frust-app-leading-functional}.

\subsection{Reconciling the finite-\texorpdfstring{$\lambda$}{lambda} and Ising
limits: the scaling variable \texorpdfstring{$\lambda T$}{lambda T}}
\label{app:frust-scaling}
 
The reduced functional \(\mathcal F\) in
Eq.~\eqref{eq:frust-app-leading-functional} describes the regime reached by
taking \(T\to0\) first at fixed \(\lambda\): the field that remains of order one
is \(\psi\), and \(\mathcal F\) knows about neither \(\lambda\) nor \(T\). The
opposite order---\(\lambda\to\infty\) first at fixed \(T\)---instead restricts
\(\phi_i\to\pm1\) and reduces the energy to the discrete form
Eq.~\eqref{eq:frust-app-1d-model}, which produces the direct FM--AF transition
reported for the discrete model in Ref.~\cite{Hu2021_2}. The two orderings
therefore define different variational problems, and the limits do not commute.
 
This non-commutativity signals that a single
combination of \(\lambda\) and \(T\) is being driven to opposite values by the
two orderings. Factoring the leading prefactor in
Eq.~\eqref{eq:frust-app-low-temperature-expansion},
\begin{equation}
E_T[\psi]
=\frac{1}{\lambda T^2}
\Big[\,\mathcal F[\psi]+(1-2\lambda)\,T\sum_i\psi_i^2\,\Big]
+\lambda N,
\label{eq:frust-app-factored}
\end{equation}
the only place the two limits differ is the coefficient \((1-2\lambda)T\) of the
mass term relative to \(\mathcal F\). Introducing the scaling variable
\begin{equation}
g\equiv\lambda T,
\label{eq:frust-app-scaling-variable}
\end{equation}
one has \((1-2\lambda)T=T-2g\). Taking \(T\to0\) at fixed \(\lambda\) sends
\(g\to0\), the mass term vanishes, and one recovers
Eq.~\eqref{eq:frust-app-leading-functional}. Taking \(\lambda\to\infty\) at
fixed \(T\) sends \(g\to\infty\) and \((1-2\lambda)T\to-\infty\), so the mass
term dominates and pins \(\psi\) (equivalently \(\phi_i=\pm1\)). The two
orderings are thus the two endpoints \(g\to0\) and \(g\to\infty\) of a single
axis.
 
The natural object interpolating between them is obtained by taking \(T\to0\)
and \(\lambda\to\infty\) \emph{jointly} at fixed \(g=\lambda T\). Then
\(T-2g\to-2g\), and the selected profile minimizes the \(g\)-dependent reduced
functional
\begin{equation}
\mathcal F_g[\psi]
=-J\sum_i\psi_i\psi_{i+1}
+\kappa J\sum_i\psi_i\psi_{i+2}
+\sum_i\psi_i^4
-2g\sum_i\psi_i^2 .
\label{eq:frust-app-Fg}
\end{equation}
The residual prefactor \(1/(gT)\) of Eq.~\eqref{eq:frust-app-Fg} in the original
energy is positive and diverges but is configuration-independent, as is the
additive \(gN/T\); neither affects the minimizer. The functional
\(\mathcal F_g\) is a Landau functional with a tunable double well: its on-site
part \(-2g\psi^2+\psi^4\) has minima at \(\psi=\pm\sqrt g\), i.e.\ exactly at
\(\phi=\psi/\sqrt{\lambda T}=\pm1\). As \(g\to0\) the well flattens to the pure
quartic and \(\mathcal F_g\to\mathcal F\); as \(g\to\infty\), writing
\(\psi_i=\sqrt g\,s_i\) gives
\begin{equation}
\frac{\mathcal F_g}{g^2}
=\frac{1}{g}\Big[
-J\!\sum_i s_i s_{i+1}
+\kappa J\!\sum_i s_i s_{i+2}\Big]
+\sum_i\big(s_i^4-2s_i^2\big),
\label{eq:frust-app-large-g}
\end{equation}
whose on-site part forces \(s_i=\pm1\) and leaves the discrete BNNNI energy at
\(O(1/g)\). The single variable \(g=\lambda T\) therefore interpolates
continuously between the soft-spin functional
Eq.~\eqref{eq:frust-app-leading-functional} and the Ising problem.
 
\emph{Branch energies.}---For \(\psi_i=A f_i\) the reduced density retains the
form Eq.~\eqref{eq:functional} with \(a_2\) shifted by the mass
term,
\begin{equation}
a_2(g)[f]=-J\langle f_i f_{i+1}\rangle+\kappa J\langle f_i f_{i+2}\rangle
-2g\langle f_i^2\rangle,
\label{eq:frust-app-a2-g}
\end{equation}
while \(a_4\) is unchanged; the amplitude minimization
Eq.~\eqref{eq:frust-app-amplitude-minimization} is therefore identical in form.
For the FM and antiphase families,
\begin{equation}
a_2^{\mathrm{FM}}(g)=-J(1-\kappa)-2g,
\qquad
a_2^{\mathrm{AF}}(g)=-\kappa J-2g,
\label{eq:frust-app-a2-fmaf}
\end{equation}
with \(a_4=1\). The mass term \(-2g\) is common to both, so \(\mathcal
F^{\mathrm{FM}}=\mathcal F^{\mathrm{AF}}\) at \(\kappa=1/2\) for every \(g\): the FM--AF boundary
is pinned at \(\kappa=1/2\) independently of \(\lambda T\), consistent both with
the \(g\to0\) continuum result and with the discrete transition of
Ref.~\cite{Hu2021_2}.
 
\emph{Fate of the IC window.}---As \(g\to\infty\), the mass term dominates
\(a_2(g)\approx-2g\langle f^2\rangle\) and
\begin{equation}
\frac{\mathcal F^{\rm min}}{N}\approx-g^2\,
\frac{\langle f^2\rangle^2}{\langle f^4\rangle}
\equiv-g^2\,R[f],
\label{eq:frust-app-shape-factor}
\end{equation}
so the selected profile maximizes the shape factor
\(R[f]=\langle f^2\rangle^2/\langle f^4\rangle\). One has \(R=1\) for a \(\pm1\)
profile (FM and AF), whereas \(R<1\) for any smooth incommensurate modulation
(\(R=2/3\) for a single cosine; additional harmonics increase \(R\) but cannot
reach unity without turning the profile into a \(\pm1\) square wave, which an
incommensurate \(q_\star\) forbids). The FM and AF branches therefore have
strictly lower leading energy than the IC branch at large \(g\), and the IC
region closes.
 
\begin{figure}[t]
    \centering
    \includegraphics[width=0.97\linewidth]{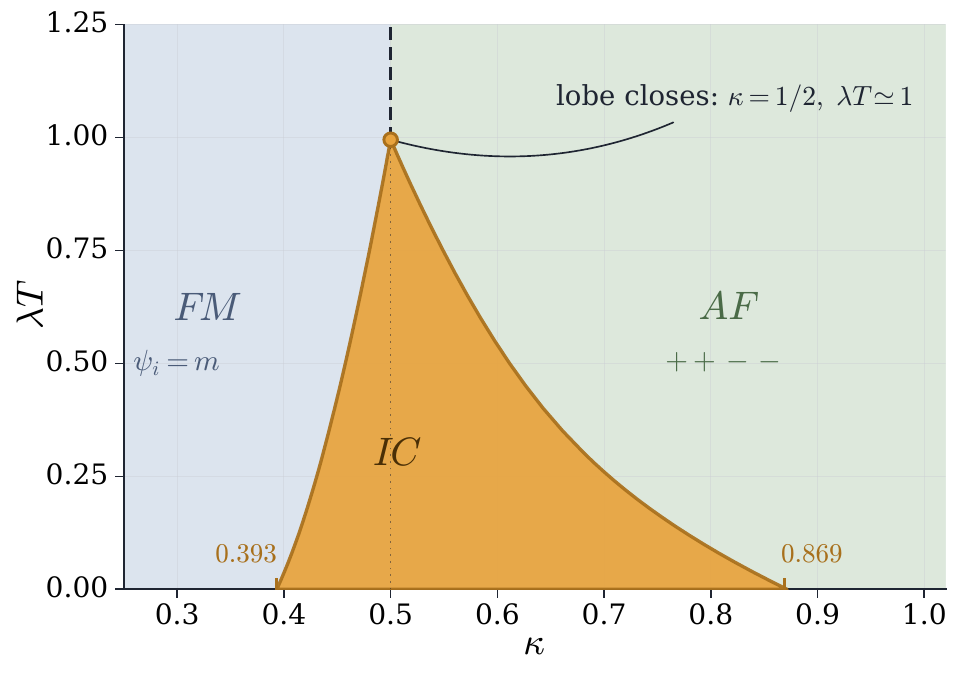}
    \caption{\small
    Restricted-variational phase diagram of the auxiliary one-dimensional model
    in the joint low-temperature, large-\(\lambda\) limit, as a function of frustration \(\kappa\) and the scaling variable
    \(\lambda T\), setting \(J=1\). The IC lobe (amber) is the region where the optimized multiharmonic
    profile has lower leading energy than both the FM (uniform) and AF
    (\(++--\)) states. It opens to \([0.393,0.869]\) as \(\lambda T\to0\),
    recovering Fig.~\ref{fig:frust-app-zero-temperature}, and closes at \(\kappa=1/2\), \(\lambda T\simeq1\), above which the FM--AF transition is
    direct (dashed line). The FM--AF boundary is pinned at \(\kappa=1/2\) for all
    \(\lambda T\). The finite-\(\lambda\) simulations of
    Fig.~\ref{fig:phase-diagram-BNNNI}, taken at low \(T\), lie near the
    \(\lambda T\to0\) edge, whereas the discrete model of Ref.~\cite{Hu2021_2}
    corresponds to the \(\lambda T\to\infty\) edge.}
    \label{fig:frust-app-lobe}
\end{figure}

The crossover is quantitative. Minimizing \(\mathcal F_g\) over the same three
families---with \(q=q_\star(\kappa)\) fixed and \(c_3,c_5,c_7,c_9\)
optimized---on a grid of \(g\) yields the IC window
\(\kappa_-(g)<\kappa<\kappa_+(g)\) shown in Fig.~\ref{fig:frust-app-lobe}. As
\(g\to0\) it opens to the interval \([0.393,0.869]\) of
Eq.~\eqref{eq:frust-app-leading-IC-window}, reproducing
Fig.~\ref{fig:frust-app-zero-temperature}; as \(g\) increases the window narrows
monotonically and pinches shut at \(\kappa=1/2\), \(\lambda T\simeq J\). At
Gaussian order the mass term is a \(q\)-independent shift, so
\(q_\star=\arccos(1/4\kappa)\) is unchanged to quadratic order; only the quartic
couplings can drift it, the same caveat noted above.
 
\emph{Interpretation.}---The dBNNNI and sBNNNI phase diagrams are the two edges of a single family controlled by \(\lambda T\). The finite-\(\lambda\) simulations of
Fig.~\ref{fig:phase-diagram-BNNNI}, performed at low \(T\), sit at small
\(g=\lambda T\) and therefore in the IC-favorable part of the plane, while the
discrete model of Ref.~\cite{Hu2021_2} corresponds to the \(g\to\infty\) edge,
where the shape penalty Eq.~\eqref{eq:frust-app-shape-factor} excludes IC and the
FM--AF transition is direct. 
 
Within its limitations---an auxiliary one-dimensional model and a
restricted multiharmonic ansatz---this calculation provides qualitative support for the
extended IC region in Fig.~\ref{fig:phase-diagram-BNNNI}, and shows explicitly
how that region connects to the direct FM--AF transition of the discrete model.

\section{Finite-size scaling of the Ising critical line}
\label{app:fss}

To determine quantitatively the location of the Ising-like critical points in the phase diagram in Fig.~\ref{fig:phase-diagram-BNNNI}, we performed a finite-size scaling analysis of the sBNNNI model at fixed frustration. For the sake of illustration, we focus here on the case
$\kappa=\kappa_x=\kappa_y=0.2$,
$\kappa_d=0$,
$\lambda=1$,
with $J=1$,
which lies on the FM-PM transition line, which belongs to the two-dimensional Ising universality class, known to be characterized by the critical exponents
$\nu=1$, and $\gamma=7/4$.

\begin{figure}[t]
    \centering
    \includegraphics[width=\linewidth]{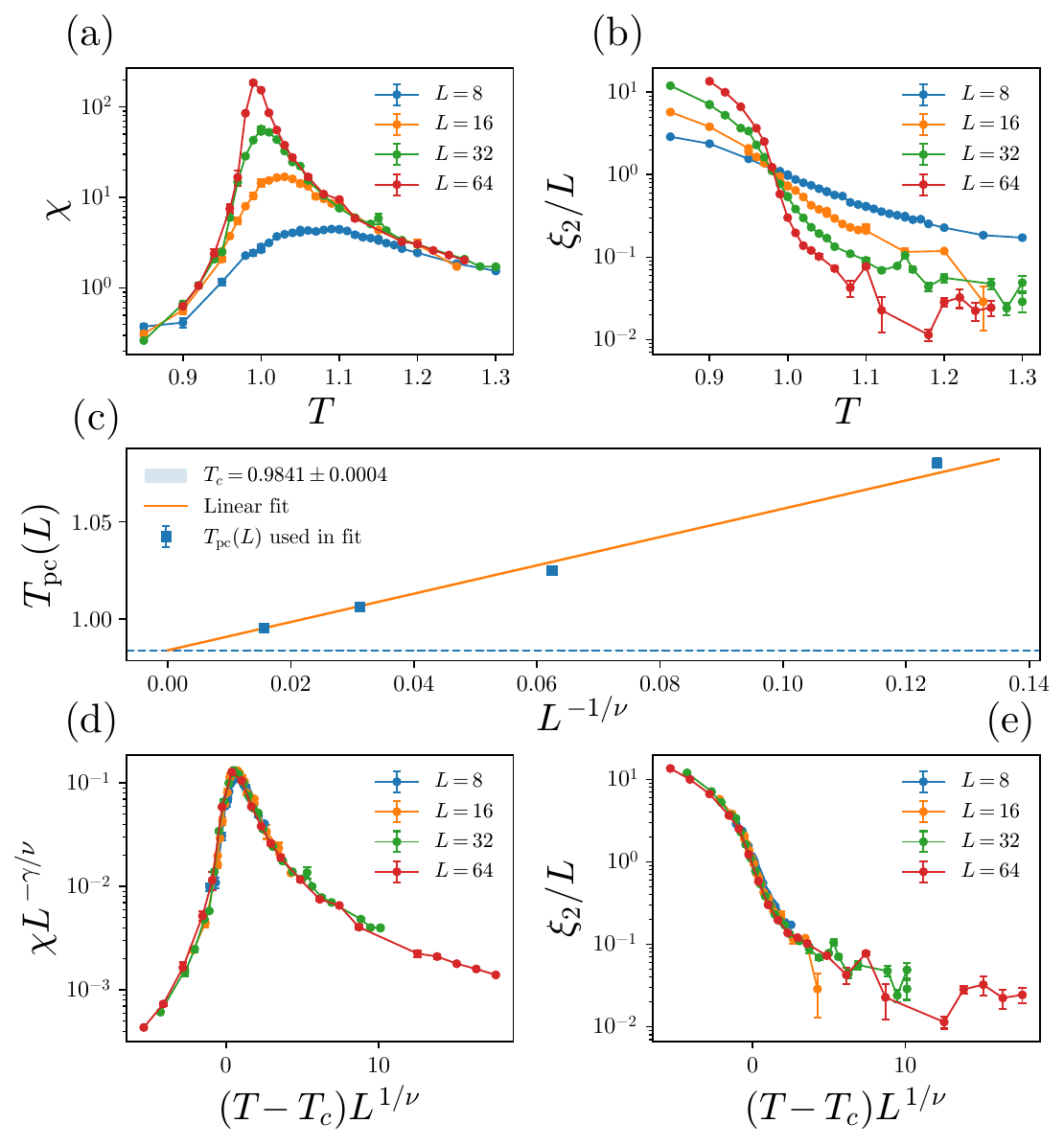}
    \caption{
    Finite-size scaling analysis of the sBNNNI model with  \(\kappa_x=\kappa_y=0.2\), \(\kappa_d=0\), \(\lambda=1\), and \(J=1\). Panels (a) and (b) show, respectively, the magnetic susceptibility \(\chi\) and the ratio \(\xi_2/L\) as functions of the temperature \(T\)  for different system sizes \(L\). Panel (c) shows the pseudocritical temperatures \(T_{\rm pc}(L)\), extracted from the maxima in panel (a), as functions of \(L^{-1/\nu}\). The solid line is the fit according to Eq.~\eqref{eq:fit-Tpc}; the dashed horizontal line indicate the resulting bulk critical temperature \(T_c\). Panels (d) and (e) show the finite-size scaling collapses of \(\chi L^{-\gamma/\nu}\) and \(\xi_2/L\), respectively, obtained using the two-dimensional Ising exponents \(\nu=1\) and \(\gamma=7/4\), together with the value of \(T_c\)
    determined in panel (c). 
    }
    \label{fig:fss_bnnni}
\end{figure}

The magnetic susceptibility $\chi$, reported in Fig.~\ref{fig:fss_bnnni}(a), was computed from the uniform magnetization \(m\)
as
$\chi(T,L) = L^2 \left( \langle m^2\rangle-\langle m\rangle^2 \right)/T$.
For each lattice size \(L\), the pseudocritical temperature \(T_{\rm pc}(L)\) corresponding to the location of the maximum of $\chi(T,L)$ with $T$  was determined by fitting the available numerical data for the peak of $\chi(T,L)$ with a quadratic polynomial. The bulk critical temperature $T_c$ was then obtained from $T_{\rm pc}(L)$ by fitting its finite-size dependence \cite{doi:10.1142/1011, PhysRevLett.28.1516} as (see Fig.~\ref{fig:fss_bnnni}(c))
\begin{equation}
T_{\rm pc}(L)
=
T_c+aL^{-1/\nu}.
\label{eq:fit-Tpc}
\end{equation}

As an additional check, we computed the second-moment correlation length
\begin{equation}
\xi_2
=
\frac{1}{2\sin(\pi/L)}
\sqrt{
\frac{S(\mathbf 0)}
     {S(\mathbf q_{\min})}
-1},
\end{equation}
where \(S(\mathbf{q})\) is the structure factor of Eq.~\eqref{eq:structure-factor} and
\(
\mathbf q_{\min}
= (2\pi/L,0)
\)
is the smallest non-zero momentum compatible with periodic boundary
conditions. 

Close to criticality, the finite-size scaling forms are
\begin{equation}
\chi(T,L)
=
L^{\gamma/\nu}
f_\chi\left((T-T_c)L^{1/\nu}\right),
\end{equation}
and
\begin{equation}
\frac{\xi_2(T,L)}{L}
=
f_\xi\left((T-T_c)L^{1/\nu}\right),
\end{equation}
where \(f_\chi\) and \(f_\xi\) are scaling functions. 
Figure~\ref{fig:fss_bnnni} shows the raw and collapsed data for $\chi$ 
and  \(\xi_2/L\). %
The quality of the collapses using the Ising exponents provides a consistency check of the
Ising character of the transition and on the value of $T_c$.

\section{Implementation of the wavelet transform}
\label{app:wavelet-conventions}

In this appendix, we briefly summarize how the wavelet transform is here implemented. At the core, we use orthogonal discrete wavelet transforms with periodic boundary
conditions. The transform is applied recursively to the microscopic field
configuration in order to separate, at each scale, slow coarse degrees of
freedom from fast wavelet degrees of freedom.

The maps \(W_j\), \(\overline W_j\), and \(\mathcal W_j\) introduced in
Sec.~\ref{sec:methods} act on two-dimensional field configurations and are used to 
construct the WCRG. In turn, they can be expressed in terms of the maps \(w_j\), \(\bar w_j\), and \(\omega_j\) specified below, which act on field configurations in one spatial dimension.
In particular, the linear maps
\(w_j\) and \(\bar w_j\) act on a field \(\varphi_{j-1}\) defined on a lattice
of size \(L_{j-1}\) and they render a coarse field \(\varphi_j\) and a wavelet
field \(\bar\varphi_j\) on a lattice of size \(L_j=L_{j-1}/2\), respectively, according to
\begin{equation}
\begin{split}
\varphi_j(x)
&=
w_j\varphi_{j-1}
=
\sum_{n=0}^{L_{j-1}-1}
g_j(n)\,
\varphi_{j-1}(2x+n),
\\[2mm]
\bar\varphi_j(x)
&=
\bar w_j\varphi_{j-1}
=
\sum_{n=0}^{L_{j-1}-1}
\bar g_j(n)\,
\varphi_{j-1}(2x+n),
\end{split}
\label{eq:app-1d-wavelet-step}
\end{equation}
where \(x=0,1,\ldots,L_j-1\). Hereafter, the arguments of the fields $\varphi_{j-1}$, $\bar \varphi_{j-1}$ and of the coefficients $g_j$ and \(\bar g_j\) are understood modulo \(L_{j-1}\).  The coefficients \(g_j\) and \(\bar g_j\) 
characterize these one-dimensional low-pass and high-pass
filters, respectively.
The low-pass component
\(\varphi_j\) contains the coarse information, while the high-pass component
\(\bar\varphi_j\) contains the fluctuations removed by the coarse graining.

The two maps define together the
wavelet step
\begin{equation}
\omega_j
=
(w_j,
\bar w_j)
,
\qquad
\omega_j\varphi_{j-1}
=
(\varphi_j,\bar\varphi_j).
\label{eq:app-1d-full-transform}
\end{equation}
The filter coefficients $g_j$ and $\bar g_j$ are chosen 
so that this map is orthogonal, i.e., 
\begin{equation}
\omega_j^{\mathrm T}\omega_j=\omega_j\omega_j^{\mathrm T}=I,
\label{eq:app-1d-wavelet-orthogonality}
\end{equation}
Orthogonality implies invertibility, with inverse given by the transpose, i.e.,
\begin{equation}
\varphi_{j-1}
=
w_j^{\mathrm T}\varphi_j
+
\bar w_j^{\mathrm T}\bar\varphi_j
\label{eq:app-1d-inverse-wavelet-matrix}
\end{equation}
which, in components, reads
\begin{equation}
\varphi_{j-1}(x)
=
\sum_{n=0}^{L_j-1}
\left[
g_j(x-2n)\,
\varphi_j(n)
+
\bar g_j(x-2n)\,
\bar\varphi_j(n)
\right],
\label{eq:app-1d-inverse-wavelet-components}
\end{equation}
where the filter indices are understood modulo \(L_{j-1}\). Accordingly, the
wavelet decomposition is invertible and no information is lost.
Orthogonality also implies that the change of variables $\varphi_{j-1} \mapsto \varphi_j, \bar\varphi_j$ has unit Jacobian, i.e.,
\begin{equation}
{\rm d}\varphi_{j-1}
=
{\rm d}\varphi_j\,{\rm d}\bar\varphi_j.
\label{eq:app-1d-wavelet-measure}
\end{equation}
The transform acting on two-dimensional configurations, used in the main text, inherits the same
property, which is used in Sec.~\ref{sec:methods} when rewriting probability
densities in terms of coarse and wavelet variables.

In addition, the filter coefficients satisfy
\begin{equation}
\sum_n g_j(n)=\sqrt{2}
\quad\mbox{and}\quad 
\sum_n \bar g_j(n)=0.
\label{eq:app-filter-sums}
\end{equation}
The first relation fixes the normalization of the coarse mode, while the
second encodes the wavelet filter having zero mean. As a result,
wavelet variables do not contribute to the uniform component of the field.
This observation is relevant for the autocorrelation analysis of the
conditional wavelet chains. At fixed \(\phi_j\), conditional sampling updates
only the wavelet variables \(\bar\phi_j\). Because the wavelet sector has zero
spatial average, these updates do not change the uniform mode of the
reconstructed field. The magnetization of the reconstructed configuration is
therefore fixed by the coarse field, up to the normalization factor, and cannot be used to
monitor the decorrelation of the conditional Monte Carlo dynamics. For this
reason, in Sec.~\ref{sec:no-csd} we use the spatial second moment of the
reconstructed field instead.

Explicit expressions for the standard wavelet filters used in this work,
including the Haar and Daubechies families, can be found in
Ref.~\cite{192463}.

We now describe the two-dimensional transform used in the simulations. The maps
\(W_j\), \(\overline W_j\), and \(\mathcal W_j\) appearing in
Sec.~\ref{sec:methods} are obtained from the one-dimensional case discussed above by
applying the filters separately along the two lattice directions. The low-low component defines the coarse
field,
\begin{equation}
\phi_j(x,y)
= 
\sum_{m,n=0}^{L_{j-1}-1}
g_j(m) g_j(n)\,
\phi_{j-1}(2x+m,2y+n),
\label{eq:app-2d-coarse}
\end{equation}
while the remaining three components define the wavelet degrees of freedom,
\begin{align}
\bar\phi_{j,1}(x,y)
&=
\sum_{m,n=0}^{L_{j-1}-1}
g_j(m) \bar g_j(n)\,
\phi_{j-1}(2x+m,2y+n),
\nonumber\\
\bar\phi_{j,2}(x,y)
&=
\sum_{m,n=0}^{L_{j-1}-1}
\bar g_j(m) g_j(n)\,
\phi_{j-1}(2x+m,2y+n),
\nonumber\\
\bar\phi_{j,3}(x,y)
&=
\sum_{m,n=0}^{L_{j-1}-1}
\bar g_j(m) \bar g_j(n)\,
\phi_{j-1}(2x+m,2y+n).
\label{eq:app-2d-wavelet-channels}
\end{align}
The three fields \(\bar\phi_{j,\mu}\), with \(\mu=1,2,3\), are referred to as
the wavelet channels at scale \(j\). We write them compactly as
\begin{equation}
\bar\phi_j
=
\left\{
\bar\phi_{j,\mu}(i)
\right\}_{i\in\Lambda_j,\ \mu=1,2,3},
\label{eq:app-wavelet-channel-notation}
\end{equation}
where \(i\) denotes a site of the coarse lattice \(\Lambda_j\) of linear size
\(L_j\). Accordingly, \(\bar\phi_j\) is a vector carrying both spatial and
channel indices.

With this notation, the two-dimensional transform used in the main text is
\begin{equation}
\mathcal W_j\phi_{j-1}
=
(\phi_j,\bar\phi_j),
\qquad
\bar\phi_j
=
\{\bar\phi_{j,\mu}\}_{\mu=1,2,3}.
\label{eq:app-2d-transform-compact}
\end{equation}

The transform preserves the number of degrees of freedom. Indeed, a field on
the lattice \(\Lambda_{j-1}\) has \(L_{j-1}^2=4L_j^2\) degrees of freedom.
After one wavelet step, these are split into \(L_j^2\) coarse variables and
\(3L_j^2\) wavelet variables, corresponding to the three wavelet channels.

Different wavelet families correspond to different choices of the filters
\(g_j(n)\) and \(\bar g_j(n)\), while the structure of the decomposition is
unchanged. As mentioned in Sec.~\ref{sec:numerical-results}, in the FM and PM
phases and close to the Ising-like critical line we use the Haar basis. In the
MP, IC, and AF regimes we use Daubechies wavelets with four vanishing moments.
The latter have wider and smoother filters and are better suited to
representing finite-wavevector modulations, whereas the Haar basis is
sufficient for uniform or long-wavelength configurations.

\section{Learning and sampling conditional energies}
\label{app:conditional-learning}

In this appendix, we describe the parameterizations used for the coarsest energy,
the conditional energies, and the conditional free-energy contributions entering
the WCRG hierarchy. We also describe how these quantities are estimated from
the original (Monte Carlo) data and how they are used either to generate synthetic configurations or to reconstruct
an explicit hierarchical energy.

At each scale \(j = 1, 2, \ldots, J\), 
the training data consist of pairs
\begin{equation}
\left\{
\phi_j^{(n)},\bar\phi_j^{(n)}
\right\}_{n=1}^{N_{\mathrm{train}}},
\label{eq:training-pairs}
\end{equation}
obtained by applying the wavelet decomposition of
App.~\ref{app:wavelet-conventions} to equilibrium configurations
\(\phi_0^{(n)}\) of the microscopic model.

\paragraph*{Coarsest energy.}

The WCRG reconstruction starts from the coarsest field \(\phi_J\), distributed
according to
\begin{equation}
p_{\theta_J}(\phi_J)
=
\frac{1}{Z_J}
\exp\left[
-
E_{\theta_J}(\phi_J)
\right].
\label{eq:app-coarsest-prob}
\end{equation}
In the present implementation the decomposition is iterated until
\(L_J=1\), so that \(\phi_J\) consists of a single degree of freedom. The
coarsest energy $E_{\theta_J}(\phi_J)$ is a one-dimensional potential $V_{\theta_J}(\phi_J)$ that is parametrized linearly, i.e.,
\begin{equation}
E_{\theta_J}(\phi_J)
=
V_{\theta_J}(\phi_J)
=
\theta_J^{\mathrm T}
\Xi_J(\phi_J).
\label{eq:app-coarsest-energy}
\end{equation}
Here
\begin{equation}
\Xi_J(u)
=
\left(
s_1(u),
s_2(u),
\ldots,
s_{N_s}(u)
\right)
\end{equation}
is a finite family of fixed one-dimensional functions, and
\begin{equation}
\theta_J
=
\left(
\theta_{J,1},
\theta_{J,2},
\ldots,
\theta_{J,N_s}
\right)
\label{eq:app-coarsest-basis}
\end{equation}
is the corresponding vector of learned coefficients.
Equivalently,
\begin{equation}
V_{\theta_J}(u)
=
\sum_{\ell=1}^{N_s}
\theta_{J,\ell}
s_\ell(u).
\label{eq:app-coarsest-potential-expansion}
\end{equation}
In this sense, the functions \(s_\ell\) are used as a sort of finite approximation basis 
for the local potential: they are fixed before training, while the optimization only
determines their linear coefficients.
Following Ref.~\cite{Brossollet2025}, we choose this finite family to be made
of translated sigmoids,
\begin{equation}
s_\ell(u)
=
\sigma\!\left(
\frac{u-u_\ell}{\Delta_\ell}
\right)
\quad\mbox{with}\quad
\sigma(z)=\frac{1}{1+e^{-z}}.
\label{eq:app-sigmoid-basis}
\end{equation}
The centers \(u_\ell\) are chosen as \(N_s\) equally spaced points covering the
empirical support of the field variable. The
widths are fixed proportionally to the spacing between consecutive centers, i.e., 
\( \Delta_\ell = (3/2)\left( u_{\ell+1}-u_\ell \right) \),
with the endpoint widths chosen by extending the same spacing convention. 
%
%
%
More generally, the wavelet decomposition discussed above, which is stopped at $L_J=1$, could be  
stopped at a finite coarse size
\(L_J>1\). In that case, \(E_{\theta_J}\) is no longer a
one-dimensional potential, but an effective energy defined on the remaining
coarse lattice. 
It could then be parametrized according to the same ``quadratic-plus-local''
structure used below for the free-energy terms, which encompasses quadratic terms involving different lattice sites and a local potential involving also higher powers of the local field.  
In the
numerical results presented here, however, we always use system sizes for which
the hierarchy reaches \(L_J=1\).

The parameters \(\theta_J\) in Eq.~\eqref{eq:app-coarsest-energy} 
are estimated from the empirical distribution of
the coarsest field using the same score-matching strategy described below for
the conditional energies.

\paragraph*{Conditional energies.}

At fixed coarse field \(\phi_j\), the parametric conditional distribution of
the wavelet variables is given by
Eq.~\eqref{eq:learned-conditional-energy-distribution}. In the implementation used in this work, the conditional energy is obtained by
first reconstructing the finer field
\(
\phi_{j-1}
=
\mathcal W_j^{-1}(\phi_j,\bar\phi_j),
\label{eq:app-reconstructed-field-conditional}
\)
and then evaluating on it the parametrized energy ansatz of Eq.~\eqref{eq:conditional-energy-ansatz}.
Thus \(\bar E_{\bar\theta_j}\) is a function of the wavelet variables
\(\bar\phi_j\) through the inverse wavelet reconstruction, while the coarse
field \(\phi_j\) is kept fixed. The model is linear in the parameters
\(\bar\theta_j\), although the basis function collected in \(\Psi_j\) are nonlinear
functions of the reconstructed field.

In the present work, the vector \(\Psi_j\) contains a translationally
invariant quadratic part and a local scalar-potential part. The quadratic part
is built from products of the reconstructed field at displaced sites. The
local part is expanded on the sigmoid functions introduced in
Eq.~\eqref{eq:app-sigmoid-basis},
\begin{equation}
\bar V_{\bar\gamma_j}(u)
=
\sum_{\ell=1}^{N_s}
\bar\gamma_{j,\ell}
s_\ell(u),
\label{eq:app-local-potential-expansion}
\end{equation}
with \(N_s=20\). The retained displacements in the quadratic part are collected
in the set \(\mathcal R_j\), which contains first- and second-neighbor
displacements on the lattice at scale \(j\), with periodic boundary conditions.

Accordingly, the complete vector of learned parameters is
\begin{equation}
\bar\theta_j
=
\left(
\left\{
\bar K_j(r)
\right\}_{r\in\mathcal R_j},
\left\{
\bar\gamma_{j,\ell}
\right\}_{\ell=1}^{N_s}
\right).
\label{eq:app-conditional-parameter-vector}
\end{equation}
The corresponding vector of basis functions is
\begin{widetext}
\begin{equation}
\Psi_j(\phi_{j-1})
=
\left(
\left\{
\frac12
\sum_{i\in\Lambda_{j-1}}
\phi_{j-1}(i)\,
\phi_{j-1}(i+r)
\right\}_{r\in\mathcal R_j},
\left\{
\sum_{i\in\Lambda_{j-1}}
s_\ell\!\left(
\phi_{j-1}(i)
\right)
\right\}_{\ell=1}^{N_s}
\right).
\label{eq:app-conditional-feature-vector}
\end{equation}
\end{widetext}
Equivalently, Eq.~\eqref{eq:conditional-energy-ansatz} can be
written explicitly as
\begin{equation}
\begin{split}
\bar E_{\bar\theta_j}(\bar\phi_j;\phi_j)
=&
\frac12
\sum_{i\in\Lambda_{j-1}}
\sum_{r\in\mathcal R_j}
\bar K_j(r)
\phi_{j-1}(i)\,
\phi_{j-1}(i+r)
\\
&+
\sum_{\ell=1}^{N_s}
\bar\gamma_{j,\ell}
\sum_{i\in\Lambda_{j-1}}
s_\ell\!\left(
\phi_{j-1}(i)
\right),
\end{split}
\label{eq:app-conditional-energy-explicit}
\end{equation}
where \(\phi_{j-1}\) is the reconstructed field.

At fixed \(\phi_j\), this ansatz defines an effective conditional energy for
the wavelet variables \(\bar\phi_j\). Additive terms depending only on
\(\phi_j\) are absorbed into the induced free energy
\(\bar F_{\bar\theta_j}^{\rm ind}\) and therefore do not affect either the
conditional probability or the score with respect to \(\bar\phi_j\).

\paragraph*{Score matching.}
In order to learn the parameters \(\bar\theta_j\), we use score matching
\cite{1046920.1088696}. At scale \(j\), the goal is to match
the score of the parametrized conditional distribution
\(p_{\bar\theta_j}(\bar\phi_j|\phi_j)\), defined in
Eq.~\eqref{eq:learned-conditional-energy-distribution}, to the score of the
exact conditional distribution \(p_j(\bar\phi_j|\phi_j)\), defined in
Eq.~\eqref{eq:conditional-prob}. The score is the gradient of the
logarithm of the conditional probability with respect to the sampled wavelet
variables \(\bar\phi_j\).

The ideal score-matching loss is therefore
\begin{equation}
\mathcal J_j(\bar\theta_j)
=
\frac{1}{2}
\left\langle
\left|
\nabla_{\bar\phi_j}
\log p_{\bar\theta_j}(\bar\phi_j|\phi_j)
-
\nabla_{\bar\phi_j}
\log p_j(\bar\phi_j|\phi_j)
\right|^2
\right\rangle_{p_{j-1}(\phi_{j-1})} .
\label{eq:app-conditional-fisher-divergence}
\end{equation}
Given
\begin{equation}
\nabla_{\bar\phi_j}
\log p_{\bar\theta_j}(\bar\phi_j|\phi_j)
=
-
\nabla_{\bar\phi_j}
\bar E_{\bar\theta_j}(\bar\phi_j;\phi_j),
\label{eq:app-conditional-score}
\end{equation}
the induced free energy\(\bar F^{\rm ind}_{\bar \theta_j}\) does not enter the score, because it depends only on
\(\phi_j\).

After integrating by parts and replacing the exact average over
\(p_{j-1}(\phi_{j-1})\) by the empirical average over the training pairs
defined in Eq.~\eqref{eq:training-pairs}, one obtains
\begin{equation}
\mathcal L_j(\bar\theta_j)
=
\left\langle
\frac{1}{2}
\left|
\nabla_{\bar\phi_j}
\bar E_{\bar\theta_j}(\bar\phi_j;\phi_j)
\right|^2
-
\Delta_{\bar\phi_j}
\bar E_{\bar\theta_j}(\bar\phi_j;\phi_j)
\right\rangle_{\mathrm{data}} .
\label{eq:app-score-matching-loss}
\end{equation}
The derivatives in Eq.~\eqref{eq:app-score-matching-loss} are taken only with
respect to the wavelet variables and are computed by automatic differentiation
of the ansatz in Eq.~\eqref{eq:app-conditional-energy-explicit}.

Using the expression of $\bar E_{\bar\theta_j}(\bar\phi_j;\phi_j)$ given in Eq.~\eqref{eq:conditional-energy-ansatz}, the loss $\mathcal L_j(\bar\theta_j)$ in Eq.~\eqref{eq:app-score-matching-loss} turns out to have a quadratic dependence on the parameters, i.e., 
\begin{equation}
\mathcal L_j(\bar\theta_j)
=
\frac{1}{2}
\bar{\theta}_j^{\mathrm T}
\overline{M}_j
\bar\theta_j
-
\bar\theta_j^{\mathrm T}
\bar b_j,
\label{eq:app-quadratic-loss}
\end{equation}
where
\begin{equation}
\overline{M}_j
=
\left\langle
\nabla_{\bar\phi_j}\Psi_j\,
\nabla_{\bar\phi_j}\Psi_j^{\mathrm T}
\right\rangle_{\mathrm{data}}
\quad\mbox{and}\quad
\bar b_j
=
\left\langle
\Delta_{\bar\phi_j}\Psi_j
\right\rangle_{\mathrm{data}}.
\label{eq:app-M-b-score-matching}
\end{equation}
The parameters that minimize the loss are then obtained by solving the regularized linear system
\begin{equation}
\left(
\overline M_j+\lambda I
\right)
\bar\theta_j
=
\bar b_j.
\label{eq:app-score-matching-linear-system}
\end{equation}
The parameter \(\lambda\)---not to be confused with the softness parameter---is used to stabilize the inversion of the
empirical matrix \(\overline M_j\): we check that moderate changes
of \(\lambda\) do not affect the observables of the synthetic configurations (obtained from the sampling of the parametrized distribution) within statistical
accuracy. 

\paragraph*{Conditinoal free energy}

The induced free energy
\(\bar F_{\bar\theta_j}^{\rm ind}(\phi_j)\) in
Eq.~\eqref{eq:learned-induced-free-energy} is fixed by the normalization of
the conditional probability. It is not required for conditional sampling,
because \(\phi_j\) is kept fixed during the sampling of \(\bar\phi_j\). However,
it is needed if one wants to reconstruct an explicit microscopic energy
functional associated with the learned WCRG hierarchy.

Following the WCRG construction of Ref.~\cite{Brossollet2025}, we approximate
the induced free energy by a finite-dimensional ansatz,
\begin{equation}
\bar F_{\bar\theta_j}^{\rm ind}(\phi_j)
\simeq
\bar F_{\alpha_j}(\phi_j)
=
\alpha_j^{\mathrm T}
\Phi_j(\phi_j),
\label{eq:app-free-energy-ansatz}
\end{equation}
where \(\Phi_j\) is a vector of functions depending only on the coarse field. In the
present work we use a quadratic-plus-local structure, analogous to Eq.~\eqref{eq:app-conditional-energy-explicit}, i.e.,
\begin{equation}
\begin{split}
\bar F_{\alpha_j}(\phi_j)
=&
\frac{1}{2}
\sum_{x\in\Lambda_j}
\sum_{r\in\mathcal R_j^{F}}
\phi_j(x)\,
\widetilde K_j(r)\,
\phi_j(x+r)\\
&+
\sum_{i\in\Lambda_j}
\widetilde V_{\widetilde\gamma_j}
\!\left(
\phi_j(i)
\right),
\end{split}
\label{eq:app-free-energy-explicit}
\end{equation}
where periodic boundary conditions are understood. 
The set
\(\mathcal R_j^{F}\) contains the displacements on the lattice which are considered when constructing the quadratic term at
scale \(j\), while \(\widetilde K_j(r)\) is a translationally invariant quadratic
kernel on the coarse lattice.

The local potential $\widetilde V_{\widetilde\gamma_j}(u)$ in the previous equation 
is expanded on the same fixed one-dimensional basis
functions as that introduced in Eq.~\eqref{eq:app-sigmoid-basis}, i.e.,
\begin{equation}
\widetilde V_{\widetilde\gamma_j}(u)
=
\sum_{\ell=1}^{N_s}
\widetilde\gamma_{j,\ell}\,
s_\ell(u).
\label{eq:app-free-energy-local-potential}
\end{equation}
Accordingly,  the parameter vector in Eq.~\eqref{eq:app-free-energy-ansatz} is given by
\begin{equation}
\alpha_j
=
\left(
\left\{
\widetilde K_j(r)
\right\}_{r\in\mathcal R_j^{F}},
\left\{
\widetilde\gamma_{j,\ell}
\right\}_{\ell=1}^{N_s}
\right).
\label{eq:app-free-energy-parameters}
\end{equation}
The corresponding vector of basis functions is
\begin{widetext}
\begin{equation}
\Phi_j(\phi_j)
=
\left(
\left\{
\frac{1}{2}
\sum_{x\in\Lambda_j}
\phi_j(x)\phi_j(x+r)
\right\}_{r\in\mathcal R_j^{F}},
\left\{
\sum_{i\in\Lambda_j}
s_\ell(\phi_j(i))
\right\}_{\ell=1}^{N_s}
\right).
\label{eq:app-free-energy-features}
\end{equation}
\end{widetext}
The parameters \(\alpha_j\) are obtained by matching the coarse-field gradient
of the ansatz \(\bar F_{\alpha_j}\) to the gradient of the induced free energy
\(\bar F_{\bar\theta_j}^{\rm ind}\). We focus on the gradient because the free
energy is defined only up to an additive constant, while its gradient is fixed
by the conditional model. Differentiating
Eq.~\eqref{eq:learned-induced-free-energy} gives
\begin{equation}
\nabla_{\phi_j}
\bar F_{\bar\theta_j}^{\rm ind}(\phi_j)
=
\left\langle
\nabla_{\phi_j}
\bar E_{\bar\theta_j}(\bar\phi_j;\phi_j)
\right\rangle_
{p_{\bar\theta_j}(\bar\phi_j|\phi_j)}.
\label{eq:app-free-energy-gradient}
\end{equation}
Formally, the average above taken over the wavelet variables sampled from the learned
conditional distribution $p_{\bar\theta_j}(\bar\phi_j|\phi_j)$ at fixed coarse field \(\phi_j\).
In the implementation used in this work, we approximate this gradient-matching
problem using the empirical wavelet--coarse pairs of Eq.~\eqref{eq:training-pairs}.
Accordingly, no additional wavelet variables are generated from
\(p_{\bar\theta_j}(\bar\phi_j|\phi_j^{(n)})\) during the free-energy fit.

The free-energy parameters are then estimated by minimizing the
gradient-matching loss
\begin{equation}
\mathcal L_j^{F}(\alpha_j)
=
\left\langle
\left|
\nabla_{\phi_j}
\bar F_{\alpha_j}(\phi_j)
-
\nabla_{\phi_j}
\bar E_{\bar\theta_j}(\bar\phi_j;\phi_j)
\right|^2
\right\rangle_{\mathrm{data}} .
\label{eq:app-free-energy-loss}
\end{equation}
Using the linear parametrizations of Eqs.~\eqref{eq:conditional-energy-ansatz} and \eqref{eq:app-free-energy-ansatz}
this becomes
\begin{equation}
\mathcal L_j^{F}(\alpha_j)
=
\left\langle
\left|
\alpha_j^{\mathrm T}
\nabla_{\phi_j}\Phi_j(\phi_j)
-
\bar\theta_j^{\mathrm T}
\nabla_{\phi_j}\Psi_j(\phi_j,\bar\phi_j)
\right|^2
\right\rangle_{\mathrm{data}} .
\label{eq:app-free-energy-loss-linear}
\end{equation}

Since \(\bar F_{\alpha_j}\) is linear in \(\alpha_j\), the minimization of
Eq.~\eqref{eq:app-free-energy-loss-linear} gives a linear regression problem.
Writing \(\Phi_{j,a}\) and \(\Psi_{j,b}\) for the components of the
vectors \(\Phi_j\) and \(\Psi_j\), we define
\begin{equation}
\left(\widetilde M_j\right)_{ab}
=
\left\langle
\nabla_{\phi_j}\Phi_{j,a}(\phi_j)
\cdot
\nabla_{\phi_j}\Phi_{j,b}(\phi_j)
\right\rangle_{\mathrm{data}},
\label{eq:app-free-energy-M}
\end{equation}
and
\begin{equation}
\left(\widetilde B_j\right)_{ab}
=
\left\langle
\nabla_{\phi_j}\Phi_{j,a}(\phi_j)
\cdot
\nabla_{\phi_j}\Psi_{j,b}(\phi_j,\bar\phi_j)
\right\rangle_{\mathrm{data}}.
\label{eq:app-free-energy-B}
\end{equation}
The fitted parameters are obtained from the regularized normal equations
\begin{equation}
\left(
\widetilde M_j+\lambda_F I
\right)
\alpha_j
=
\widetilde B_j
\bar\theta_j,
\label{eq:app-free-energy-normal-equation}
\end{equation}
or equivalently
\begin{equation}
\alpha_j
=
\left(
\widetilde M_j+\lambda_F I
\right)^{-1}
\widetilde B_j
\bar\theta_j.
\label{eq:app-free-energy-linear-system}
\end{equation}
The parameter \(\lambda_F\) regularizes the empirical regression matrix
\(\widetilde M_j\), in the same way as \(\lambda\) regularizes the
score-matching regression for the conditional energies. 

The dependence of \(\alpha_j\) on \(\bar\theta_j\) simply reflects the fact
that the target gradient in Eq.~\eqref{eq:app-free-energy-loss} is computed
from the conditional energy model learned in the previous step. The free-energy
fit therefore does not introduce an independent conditional model; it provides
an explicit parametrization of the normalization induced by
\(\bar E_{\bar\theta_j}\).

\paragraph*{Conditional sampling.}

Once \(\bar\theta_j\) has been learned, the conditional distribution \(\bar p_{\bar\theta_j}(\bar\phi_j|\phi_j)\) is sampled using a standard local Metropolis Monte Carlo algorithm.
A single wavelet coefficient is updated according to
\begin{equation}
\bar\phi_{j,a}
\longrightarrow
\bar\phi_{j,a}'
=
\bar\phi_{j,a}
+
\eta,
\label{eq:app-metropolis-proposal}
\end{equation}
where \(a\) labels both the spatial position and the wavelet channel and $\eta$ is a  random variable uniformly distributed over the interval $[-\delta,\delta]$.
The move is then accepted with probability
\begin{equation}
P_{\mathrm{acc}}
=
\min\left[
1,
e^{-\Delta \bar E_{\bar\theta_j}}
\right],
\label{eq:app-metropolis-acceptance}
\end{equation}
with
\(
\Delta \bar E_{\bar\theta_j}
\) is given by  Eq.~\eqref{eq:delta-barE}.
The proposal amplitude \(\delta\) is tuned, during a short warm-up stage, so as to
obtain an acceptance rate of roughly~\(50\%\).

The same local Metropolis scheme is used to generate the original Monte Carlo
configurations of the microscopic model. In that case, the proposed
update is applied to a single microscopic field variable \(\phi_0(i)\), and the
acceptance probability is computed from the corresponding change
\(\Delta E_0\) of the microscopic energy in Eq.~\eqref{eq:Hsoft}.

The choice of local conditional updates should be contrasted with global
Langevin-based proposals, such as MALA
\cite{10.1214/11-AAP828,10.1111/1467-9868.00123}, which have also been used in
related learned multiscale sampling contexts \cite{pmlr-v202-guth23a}. In such
methods, the update of the full set of variables is proposed at once, then accepted or rejected. Global proposals are naturally parallelizable---in contrast to local
Metropolis sweeps---but their acceptance and decorrelation properties typically
degrade as the number of degrees of freedom increases. For example, optimal
scaling results for MALA give a characteristic scaling of order \(N^{1/3}\)
decorrelation steps for a target distribution of \(N\) variables. In two spatial
dimensions, where \(N=L^2\), this corresponds to \(L^{2/3}\). Local conditional Metropolis updates also
provide a direct measure the decorrelation time of the conditional chain at each reconstruction
scale (in units of MC sweeps).

\paragraph*{Configuration and energy reconstruction.}
In order to generate a microscopic configuration, we first sample the coarsest field
\(\phi_J\) from \(p_{\theta_J}(\phi_J)\). Then, for
\(j=J,J-1,\ldots,1\), we sample the wavelet variables
\(\bar\phi_j\) from \(p_{\bar\theta_j}(\bar\phi_j|\phi_j)\) and reconstruct $\phi_{j-1}$ with the inverse filter of Eq.~\eqref{eq:W-1-filter}.
For this generative procedure, only the knowledge of \(E_{\theta_J}\) and of the conditional
energies \(\bar E_{\bar\theta_j}\) is required.

The fitted free-energy terms \(\bar F_{\alpha_j}\) are instead used to
construct an explicit hierarchical energy for the learned WCRG model, according to Eq.~\eqref{eq:wcrg-hierarchical-energy}.
Equivalently,
\begin{equation}
\begin{split}
p_0^{\rm WCRG}(\phi_0)
&\propto
\exp\left[
-
E_0^{\rm WCRG}(\phi_0)
\right]\\
&=
p_{\theta_J}(\phi_J)
\prod_{j=1}^{J}
p_{\bar\theta_j}(\bar\phi_j|\phi_j).
\label{eq:app-wcrg-hierarchical-measure}
\end{split}
\end{equation}

The energy \(E_0^{\rm WCRG}\) is the microscopic energy induced by the learned
hierarchical model. It is not, in general, identical to the original
microscopic energy \(E_0\), because the conditional energies and the
free-energy contributions are both represented in terms of finite-dimensional
ans\"atze.

\input{references.bbl}

\end{document}

%% file: references.bbl
%